\documentclass[12pt]{article}
\usepackage[dvips]{graphicx}
\usepackage{latexsym}
\usepackage[abs]{overpic}
\usepackage{wrapfig}
\usepackage{amssymb}
\usepackage{amsmath}
\usepackage{lscape}
\usepackage{bbm}
\usepackage{color}

\input epsf
\newcommand{\skyp}[1]{}

\def\Z {\bb{Z}}

\def\rem#1{}

\def\b{{\mathfrak{b}}}
\def\C{{\mathbb{C}}} 
\def\Z{{\mathbb{Z}}} 
\def\D{{\mathcal{D}}} 

\def\lam{{\lambda}}
\def\Tr{{\mathrm{Tr}}}

\def\Ad{{\mathrm{Ad}}}
\def\Det{{\mathrm{Det}}}

\def\oo{{\ddot{\text{o}}}}

\newcommand{\bel}{\begin{eqnarray}}
\newcommand{\ee}{\end{eqnarray}}

\newtheorem{theorem}{Theorem}[section]

\makeatletter
\@addtoreset{equation}{section}
\makeatother
\begin{document}

\begin{titlepage}

\bigskip
\hfill\vbox{\baselineskip12pt
\hbox{KEK-TH-1658}
\hbox{RUP-13-8}
}
\bigskip\bigskip\bigskip\bigskip

\begin{center}
\LARGE{\sf G/G gauged WZW-matter model, Bethe Ansatz 
for q-boson model and Commutative Frobenius algebra}
\end{center}

\bigskip
\bigskip

\bigskip
\bigskip
\centerline{ \large Satoshi Okuda$^1$ and Yutaka Yoshida,$^2$}
\bigskip
\medskip
\centerline{$^1$\it Department of Physics, Rikkyo University}
\centerline{\it Toshima, Tokyo 171-8501, Japan, \ \ }
\centerline{ okudas@rikkyo.ac.jp}
\bigskip
\centerline{$^2$\it High Energy Accelerator Research Organization (KEK)}
\centerline{\it Tsukuba, Ibaraki 305-0801, Japan}
\centerline{ yyoshida@post.kek.jp}
\bigskip
\bigskip

\bigskip
\bigskip

\begin{abstract}
We investigate the correspondence between two dimensional topological gauge theories and quantum integrable systems discovered by Moore, Nekrasov, Shatashvili.
This correspondence means that the hidden quantum integrable structure exists in the topological gauge theories.
We showed the correspondence between the G/G gauged WZW model and the phase model in JHEP 11 (2012) 146 (arXiv:1209.3800).
In this paper, we study a one-parameter deformation for this correspondence and
show that the G/G gauged WZW model coupled to additional matters corresponds to the $q$-boson model.
Furthermore, we investigate this correspondence from the viewpoint of  the commutative Frobenius algebra, the axiom of the two dimensional topological quantum field theory.
\end{abstract}

\end{titlepage}

\newpage
\baselineskip=18pt

\tableofcontents
\section{Introduction}
We investigate the correspondence between topological field theories and quantum integrable systems, discovered by Moore, Nekrasov and Shatashvili \cite{Moore:1997dj}.
They applied the cohomological localization method to the topological Yang-Mills-Higgs model and then discovered that its localized configurations coincide with the Bethe Ansatz equations in the non-linear Schr$\oo$dinger model.
Later, Gerasimov and Shatashvili revealed that the partition function of the topological Yang-Mills-Higgs model is related to the norms of the wave functions in the non-linear Shcr$\oo$dinger model \cite{Gerasimov:2006zt}.
From this fact, the topological Yang-Mills-Higgs model corresponds to the non-linear Schr$\oo$dinger model.
We call the correspondence like this as the Gauge/Bethe correspondence through this paper.
This  correspondence implies that a special topological gauge theory has a hidden quantum integrable structure.

In the previous paper \cite{Okuda:2012nx}, we showed a correspondence between the $G/G$ gauged Wess-Zumino-Witten (WZW) model  on a genus-$h$ Riemann surface $\Sigma_h$ 
and the phase model, which is a  quantum integrable field theory on a one-dimensional lattice and a strongly correlated boson system, first introduced by \cite{Bogoliubov:1993}.
In particular, we showed that its localized configurations and the partition function coincide with the Bethe Ansatz equations and a summation of all  the norms between the eigenstates of the Hamiltonian in the phase model, respectively.
Furthermore, the $G/G$ gauged WZW model is equivalent to the Chern-Simons theory with the gauge group $G$ on $S^1 \times \Sigma_h$
because the partition function of the $G/G$ gauged WZW model coincides with that of the Chern-Simons theory \cite{Witten:1988hf, Blau:1993tv}.
Therefore, we found that the Chern-Simons theory on $ S^1 \times \Sigma_h$ also corresponds to the phase model.

The Gauge/Bethe correspondence is realized for not only  topological gauge theories but also vacua in  a supersymmetric gauge theory.
Nekrasov and Shatashvili discovered that coulomb branch in a supersymmetric gauge theory corresponds to a certain integrable system.
For example, they found that the effective twisted superpotential in an $\mathcal{N}=(2,2)$ supersymmetric gauge theory in two dimensions 
coincides  with the Yang-Yang function for the XXX model \cite{Nekrasov:2009ui, Nekrasov:2009uh}. 
This correspondence is deeply related to the Gauge/Bethe correspondence between topological field theories and  quantum integrable systems.
This is natural because the vacua of the supersymmetric gauge theory transfer to  physical states in the topological field theory through a topological twist.
 Although it is known that various supersymmetric or topological gauge theories correspond to certain quantum integrable systems,
the underlying mathematical principle of the Gauge/Bethe correspondence is not clear up to now.

Our purpose is to construct  a one-parameter deformation of the Gauge/Bethe correspondence 
between the $G/G$ gauged WZW model and the phase model,
and to investigate the underlying mathematical principle  of the Gauge/Bethe correspondence in our case.
It is known that the phase model can be realized by the strong coupling limit of the $q$-boson model, a quantum integrable field theory on a one-dimensional lattice \cite{Bogoliubov:1993, Bogoliubov:1997}.
Therefore the $q$-boson model can  be regarded as the one-parameter deformation of the phase model.
From the viewpoint of the Gauge/Bethe correspondence,
we expect that there exists a one-parameter deformation of the $G/G$ gauged WZW model corresponding to the $q$-boson model.
Such a model actually exists  and is the $G/G$ gauged WZW model coupled to additional scalar matters.  
We call this model as the $G/G$ gauged WZW-matter model.
We will establish a new correspondence between the $G/G$ gauged WZW-matter model and the $q$-boson model by utilizing the cohomological localization method in this paper.

We also study the Gauged WZW-matter model/$q$-boson model correspondence from the viewpoint of the axiomatic system of the  topological quantum field theory (TQFT) given by Atiyah \cite{Atiyah:1989vu} and Segal \cite{Segal:2002ei} in order to investigate the  underlying mathematical principle of this correspondence.
In particular, it is well known that the category of commutative Frobenius algebras 
is categorical equivalent to that of  two dimensional TQFTs, e.g. \cite{Dijkgraaf:1997ip, Dijkgraaf}.
Recently, Korff constructed a new commutative Frobenius algebra from the $q$-boson model \cite{Korff:2013rsa} as a one-parameter deformation of the Verlinde algebra in the Wess-Zumino-Witten model constructed from the phase model \cite{Korff:2010}.
Thus, it  is natural to think that there exists a relation between the $SU(N)/SU(N)$ gauged WZW-matter model and the TQFT equivalent to this commutative Frobenius algebra, as with the relation between the $SU(N)/SU(N)$ gauged WZW model and the Verlinde algebra \cite{Okuda:2012nx}.
We will show equivalence between the $SU(N)/SU(N)$ gauged WZW-matter model and the topological field theory constructed by Korff.

This paper is organized as follows.
In section \ref{sec:q-boson}, we review the $q$-boson model and the algebraic Bethe Ansatz for this model.
In particular, we give a determinant formula for norms between the eigenstates of the Hamiltonian in the $q$-boson model.
This norm will become one of the most important quantities when we consider the Gauge/Bethe correspondence.
In section \ref{sec:GWZWM}, we investigate the Gauge/Bethe correspondence between  the $G/G$ gauged WZW-matter model and the $q$-boson model.
In order to establish the correspondence, we construct the $G/G$ gauged WZW-matter model in section \ref{subsec:GWZWM}.
Later, we apply the cohomological localization method to this model in the case of $G=U(N)$, and derive its partition function in section \ref{subsec:GWZWM localization}.
In section \ref{subsec:GWZWM numerical}, we evaluate numerically the partition function for several cases with different $N$ and the level $k$.
In section \ref{subsec:GWZWM Gauge/Bethe}, we establish the Gauge/Bethe correspondence between the $SU(N)/SU(N)$ or $U(N)/U(N)$ gauged WZW-matter model and the $q$-boson model.
In section \ref{subsec:GWZWM Axiom}, we study the correspondence between the $SU(N)/SU(N)$ gauged WZW-matter model and the $q$-boson model from the viewpoint of the axiomatic system of the TQFT
and investigate relations with the TQFT constructed by Korff. 
In section \ref{sec:Correlation}, we extend the Gauge/Bethe correspondence for the partition function to that for correlation functions. 
We show the correspondence between the correlation functions of gauge invariant  BRST-closed operators in the $SU(N)/SU(N)$ gauged WZW-matter model and
the expectation values of conserved charges in the $q$-boson model. 
The final section is devoted to the summary and the discussion.


\section{$q$-boson model}
\label{sec:q-boson}

In this section, we introduce the $q$-boson model and apply the algebraic Bethe Ansatz to this model.
The $q$-boson model is a quantum integrable field theory on a one-dimensional lattice
and is regarded as the $q$-deformation of the free boson system on the lattice.
Also, this model becomes the phase model in the strong coupling limit $q\rightarrow 0$.
See  \cite{Bogoliubov:1993, Bogoliubov:1997,  Korff:2013rsa, Korff:2010, Korepin:book, Bethe:1931hc} for the $q$-boson model  and the algebraic Bethe Ansatz method in details. 
%

\subsection{$q$-boson model}
\label{subsec:q-boson}

Let us define the $q$-boson model.
First, consider the operators  $\{q^{\pm \hat{N}},\beta,\beta^{\dagger}\}$ which satisfy  the $q$-boson algebra (or the $q$-oscillator algebra) ${\cal H}_q$: 
\begin{eqnarray}
&&q^{\hat{N}}q^{-\hat{N}} = q^{-\hat{N}}q^{\hat{N}} =1,~~q^{\hat{N}}{\beta} = \beta q^{\hat{N}-1},~~q^{\hat{N}} \beta^{\dagger} = \beta^{\dagger} q^{\hat{N}+1},
\nonumber\\
&&\beta\beta^{\dagger} - \beta^{\dagger}\beta = (1-q^2)q^{2\hat{N}},~~\beta\beta^{\dagger} - q^2\beta^{\dagger}\beta=1-q^2
\label{q-boson algebra}
\end{eqnarray}
where $q^{\pm \hat{N}}$ denotes generators and $q^{\pm p\hat{N}+x}$ is the abbreviation of $(q^{\pm \hat{N}})^p q^x$.
The parameter $q$ is a generic real $c$-number and $0\le q <1$.
From this algebra, we find that the operators $\hat{N}$, $\beta$ and $\beta^{\dagger}$ serve as the number operator, the annihilation operator and the creation operator, respectively.

Next, we construct a Fock space ${\cal F}$ for the $q$-boson algebra given by (\ref{q-boson algebra}).
The Fock space is constructed as 
\begin{eqnarray}
q^{\hat{N}}|m\rangle = q^m|m\rangle,
\quad 
\beta^{\dagger}|m\rangle = (1-q^{2m+2})|m+1\rangle,
\quad
\beta|m\rangle = | m-1\rangle.
\end{eqnarray}
The basis is given by the set $\{|m\rangle = (\beta^{\dagger})^m/(q^2)_m|0\rangle ~|~ m\in \Z_{\ge 0}\}$
where $(x)_m$ is $(x)_m = \prod_{i=0}^{m-1}(1 - x^{i+1})$.
Also, $|0\rangle$ is defined as a state which is annihilated by acting on the annihilation operator $\beta$.

In order to define the Hamiltonian of the $q$-boson model, we generalize the $q$-boson algebra and the Fock space to their $L$-fold tensor product.
We denote the operators as $\{\beta_i,\beta_i^{\dagger},q^{\hat{N}_i}\}_{i=1,\cdots,L}$
and define the $L$-fold tensor product ${\cal H}_q^{\otimes L}$ of the $q$-boson algebra (\ref{q-boson algebra}) as
\begin{eqnarray}
\beta_i\beta_j - \beta_j \beta_i &=& \beta_i^{\dagger}\beta^{\dagger}_j - \beta^{\dagger}_j \beta^{\dagger}_i 
=q^{\hat{N}_i}q^{\hat{N}_j} - q^{\hat{N}_j} q^{\hat{N}_i}=0, \nonumber\\
q^{\hat{N}_i}\beta_j &=& \beta_j q^{\hat{N}_i - \delta_{i j}},
\quad q^{\hat{N}_i}\beta_j^{\dagger} = \beta_j^{\dagger} q^{\hat{N}_i +\delta_{i j}}, \nonumber\\
\beta_i\beta_j^{\dagger} - \beta_j^{\dagger} \beta_i &=& \delta_{i j}(1-q^2)q^{2\hat{N}_i},
\quad \beta_i\beta_i^{\dagger} - q^2 \beta_i^{\dagger} \beta_i = (1-q^2).
\label{q-boson algebra tensor}
\end{eqnarray}
Also, we  can define the $L$-fold tensor product of the Fock space ${\cal F}^{\otimes L}$ just like the case of $L=1$.
The basis of ${\cal F}^{\otimes L}$ is given by the set
$\{|m_1,\cdots,m_ L\rangle = |m_1\rangle\otimes\cdots\otimes|m_L\rangle ~| m_i\in \Z_{\ge 0}\}$.

By using these relations, we  define the Hamiltonian of the $q$-boson model which belongs to ${\cal H}_q^{\otimes L}$ and acts on ${\cal F}^{\otimes L}$.
The Hamiltonian of the $q$-boson model with the periodic boundary condition $L+1 \equiv 1$ and with the total site number $L$ is as follows:
\begin{eqnarray}
H = -\frac{1}{2}\sum_{j=1}^L \left(\beta_j\beta^{\dagger}_{j+1} + \beta^{\dagger}_j\beta_{j+1}\right)
\label{q-boson Hamiltonian}
\end{eqnarray}
where  the lattice spacing is $1$ and the index of the operators $j$ labels a site of the lattice.

In order to understand the properties of the $q$-boson model, 
we consider relations between the $q$-boson algebra (\ref{q-boson algebra tensor}) and the harmonic oscillator algebra
\begin{eqnarray}
[\hat{N}_i, a_j] = - a_i \delta_{i j},\quad
[\hat{N}_i, a^{\dagger}_j] = a^{\dagger}_i \delta_{i j},\quad
[a_i, a^{\dagger}_j] = \delta_{i j}.
\end{eqnarray}
The operators obeying the $q$-boson algebra are represented by the operators $\{a_i, a_i^{\dagger}, N_i\}$ obeying the harmonic oscillator algebra as follows:
\begin{eqnarray}
\beta_i = \sqrt{\frac{1 -q^{2(\hat{N}_i+1)}}{1 + \hat{N}_i}} a_i,\quad
\beta_i^{\dagger} = a^{\dagger}_i \sqrt{\frac{1 -q^{2(\hat{N}_i+1)}}{1 + \hat{N}_i}}
\label{substitution q-boson}
\end{eqnarray}
where these are defined as a formal power series.

We rewrite the Hamiltonian (\ref{q-boson Hamiltonian}) by using the substitution (\ref{substitution q-boson}) as
\begin{eqnarray}
H 
&=& \sum_{j=1}^L \left(\sqrt{\frac{1 -q^{2(\hat{N}_{j}+1)}}{1 + \hat{N}_{j}}}a_j a^{\dagger}_{j+1} \sqrt{\frac{1 -q^{2(\hat{N}_{j+1}+1)}}{1 + \hat{N}_{j+1}}}\right.\nonumber\\
&&\hspace{4cm}\left. +a^{\dagger}_j \sqrt{\frac{1 -q^{2(\hat{N}_j+1)}}{1 + \hat{N}_j}}\sqrt{ \frac{1 -q^{2(\hat{N}_{j+1}+1)}}{1 + \hat{N}_{j+1}}} a_{j+1} \right).
\end{eqnarray}
Here, $q$ serves as a coupling constant of the $q$-boson model.
When we expand the Hamiltonian in terms of the coupling constant, infinite interaction terms  appear in front of the hopping term.
Therefore we find that the $q$-boson model is  the strongly interacting system
and the quantum field theory with non-local interactions on the lattice.

Also, we find that the $q$-boson algebra and the Hamiltonian of the $q$-boson model 
reduce to  those of the free boson at the leading order of $\xi$, once we set $q = e^{\xi}$ and expand it around $\xi =0$.
Thus  the $q$-boson is regarded as the $q$-deformation of the usual free boson  in  the weak coupling $q \sim 1 ~(\xi \sim 0)$.
On the other hand, the $q$-boson model becomes the phase model in the strong coupling limit $q\rightarrow 0 ~(\xi \rightarrow -\infty)$.
There also exists a continuum limit because the $q$-boson model is a field theory on the lattice.
In this limit, the $q$-boson model becomes the non-linear Schr$\oo$dinger model.

\subsection{Algebraic Bethe Ansatz for the $q$-boson model}
\label{subsec:q-boson ABA}

In this subsection, we apply the algebraic Bethe Ansatz to the $q$-boson model.
In particular, we construct the eigenvalues and the eigenstates of the Hamiltonian and give Bethe Ansatz equations.
Furthermore, we give a determinant formula for norms between the eigenstates of the Hamiltonian.

In order to apply the algebraic Bethe Ansatz method to the $q$-boson model, we first define an L-matrix  and an R-matrix which satisfy the Yang-Baxter equation:
\begin{eqnarray}
R(\mu,\nu) ({\cal L}(\mu)\otimes {\cal L}(\nu)) =  ({\cal L}(\nu)\otimes {\cal L}(\mu)) R(\mu,\nu).
\label{Yang-Baxter}
\end{eqnarray}
The L-matrix of the $q$-boson model at a site $n ~(n=1,\cdots,L)$ is  a matrix in an auxiliary space $\C^2$ and is defined by
\begin{eqnarray}
{\cal L}_n(\mu) = \left(
\begin{array}{cc}
1&~~\mu\beta^{\dagger}_n\\
\beta_n&~~\mu
\end{array}
\right)
\quad
\in {\rm End}[\C^2(\mu)]\otimes {\cal H}_q
\end{eqnarray}
where $\mu \in \C$ is a spectral parameter. 
$\beta_n$ and $\beta^{\dagger}_n$ obey the $q$-boson algebra (\ref{q-boson algebra tensor}).
Also, the R-matrix is defined by
\begin{eqnarray}
R(\mu,\nu) = \left(
\begin{array}{cccc}
f(\nu,\mu)&0&0&0\\
0&g(\mu,\nu)&1&0\\
0&t&-g(\nu,\mu)&0\\
0&0&0&f(\nu,\mu)
\end{array}
\right)
\quad
\in {\rm End}[\C^2(\mu)\otimes\C^2(\nu)]\label{R matrix}
\end{eqnarray}
where
\begin{eqnarray}
f(\mu,\nu) =\frac{\mu t-\nu}{\mu-\nu},\quad
g(\mu,\nu)=\frac{(1-t)\nu}{\mu-\nu}\quad
\mathrm{and}\quad t=q^2.
\end{eqnarray}

Next, we define the monodromy matrix as
\begin{eqnarray}
T(\mu) = {\cal L}_L(\mu){\cal L}_{L-1}(\mu)\cdots {\cal L}_1(\mu)
=\left(
  \begin{array}{cc}
    A(\mu)   & B(\mu)   \\
     C(\mu)   &  D(\mu)    \\
  \end{array}
\right).
\label{monodromy matrix}
\end{eqnarray}
Then, we  can show  the following relation from the Yang-Baxter equation (\ref{Yang-Baxter}): 
\begin{eqnarray}
R(\mu,\nu) (T(\mu)\otimes T(\nu)) =  (T(\nu)\otimes T(\mu)) R(\mu,\nu).
\label{TTR relation}
\end{eqnarray}
From this formula, we can derive 16 commutation relations for the elements of  the monodromy matrix,
$A(\mu),B(\mu),C(\mu),D(\mu)$. For example,
\begin{eqnarray}
[{\cal O}(\mu), {\cal O}(\nu)] &=& 0, \quad \text{for}~{\cal O} = A, B, C, D, \label{commutationAA}\\
A(\mu)B(\nu) &=& f(\mu,\nu) B(\nu)A(\mu) + g(\nu,\mu)  B(\mu)A(\nu), \label{commutationAB}\\
D(\mu)B(\nu) &=& f(\nu,\mu) B(\nu)D(\mu) - g(\mu,\nu) B(\mu)D(\nu),\label{commutationDB}\\
C(\mu)A(\nu) &=& f(\nu,\mu) A(\nu)C(\mu) - g(\mu,\nu) A(\mu)C(\nu),\label{commutationCA}\\
C(\mu)D(\nu) &=& f(\mu,\nu) D(\nu)C(\mu) + g(\mu,\nu) D(\mu)C(\nu),\label{commutationCD}\\
C(\mu)B(\nu) - t B(\nu)C(\mu)
&=& g(\mu,\nu)(D(\mu)A(\nu)-D(\nu)A(\mu))\label{commutationCB1}\\
&=& g(\mu,\nu)(A(\nu)D(\mu) - A(\mu)D(\nu)).\label{commutationCB2}
\end{eqnarray}
The transfer matrix is defined by taking trace of the monodromy matrix with respect to the auxiliary space:
\begin{eqnarray}
\tau(\mu) = \mathrm{tr}T(\mu) = A(\mu) +  D(\mu).
\end{eqnarray}
We can show that the transfer matrices with the different spectral parameters commute
by  taking trace  of the both sides of (\ref{TTR relation}) with respect to the auxiliary space ${\rm End}[\C^2(\mu)\otimes \C^2(\nu)]$:
\begin{eqnarray}
[\tau(\mu),\tau(\nu)] =0.
\label{commute}
\end{eqnarray}
By expanding the transfer matrix
 as a power series $\tau(\mu) = \sum_{a=0}^{L}H_a \mu^a$ and substituting it to (\ref{commute}),  we show that all the operators $\{H_0, H_1, \cdots, H_L\}$  commute.
Therefore, the transfer matrix can be regarded as a generating function of the conserved charges.
Note that $H_0$ and $H_L$ are not conserved charges because of $H_0 = H_L =1$.
Also, the Hamiltonian of the $q$-boson model (\ref{q-boson Hamiltonian}) is expressed via the conserved charges as
\begin{eqnarray}
H = - \frac{1}{2}(H_1 + H_{L-1}).
\end{eqnarray}
Putting together the total particle number operator and $\{ H_1,\cdots, H_{L-1}\}$,
we find that the $q$-boson model possesses as many commuting conserved charges  as the degree of freedom of the system.
Therefore, the $q$-boson model is a quantum integrable system.

From now on, let us construct the eigenvalues and the eigenvectors of the transfer matrix.
Since $C(\mu)$  and $B(\mu)$ are an annihilation operator and an creation operator, respectively, 
the vacuum state $|0\rangle$ and its dual vacuum state $\langle 0|$ satisfy $C(\mu)|0\rangle=0$ and $\langle 0|B(\mu) = 0$.
Also, the eigenvalues of operators $A(\mu)$ and $D(\mu)$  on the vacuum state are $|0\rangle$ $a(\mu)=1$ and $d(\mu) = {\mu}^L$ , respectively.

Suppose that  a state $\prod_{j=1}^M B(\lambda_j)|0\rangle$ is the eigenstate of the transfer matrix:
\begin{eqnarray}
\tau(\mu)\prod_{j=1}^M B(\lambda_j)|0\rangle 
= \Lambda(\mu,\{\lambda\})\prod_{j=1}^M B(\lambda_j)|0\rangle
\label{remainder term}
\end{eqnarray}
where the eigenvalue of the transfer matrix $\Lambda(\mu,\{\lambda\})$:
\begin{eqnarray}
\Lambda(\mu,\{\lambda\})=a(\mu)\prod_{j=1}^M f(\mu,\lambda_j) +d(\mu)\prod_{j=1}^M f(\lambda_j,\mu).
\label{transfer matrix eigenvalue}
\end{eqnarray}
Then,  the spectral parameters $\{\lambda_1,\cdots,\lambda_M\}$ must  satisfy  the Bethe Ansatz equations
\begin{eqnarray}
a(\lambda_j)\prod_{\substack{k=1\\k\neq j}}^M f(\lambda_j,\lambda_k)
=d(\lambda_j)\prod_{\substack{k=1\\k\neq j}}^M f(\lambda_k,\lambda_j)
\quad\quad \mathrm{for}\quad j=1,\cdots,M.
\label{Bethe equation q-boson 1}
\end{eqnarray}
The Bethe Ansatz equations concretely are
\begin{eqnarray}
\lambda_j^{L} = \prod_{\substack{k=1\\k\neq j}}^M\frac{\lambda_j t -\lambda_k}{\lambda_j - \lambda_k t} \quad\quad \mathrm{for}\quad j=1,\cdots,M.
\label{Bethe equation}
\end{eqnarray}
Note that the Bethe roots assign the ground state or excited states in the $q$-boson model.
Also, we call the state $\prod_{j=1}^M B(\lambda_j)|0\rangle$ with the spectral parameters $\{\lambda_j\}$ which satisfy the Bethe Ansatz equations, as Bethe vector.

Here, we summarize the several properties of the Bethe Ansatz equations.
For  convenience, we change a parameterization of the Bethe roots as $\lambda_j = e^{2\pi i x_j}$ for $j=1,\cdots,M$ and of the coupling constant $t$ as $t=e^{-2\pi \eta}$ where $\eta>0$ because of $0\le t < 1$. 
Then, the Bethe Ansatz equations become
\begin{eqnarray}
e^{2\pi i L x_j} = \prod_{\substack{k=1\\k\neq j}}^M \frac{\sin[\pi(x_j-x_k + i\eta)]}{\sin[\pi(x_j-x_k - i\eta)]} \quad\quad \mathrm{for}~j=1,\cdots,M.
\label{exp Bethe equation}
\end{eqnarray}
From this equations, we can show that the Bethe roots $\{x_1,\cdots,x_M\}$ are real numbers by using a similar manner with the Bose gas model \cite{Yang:1968rm, Korepin:book}.
The logarithmic form of (\ref{exp Bethe equation}) is
\begin{eqnarray}
2\pi i L x_j =  2\pi i I_j + \sum_{k=1}^M \log\frac{\sin[\pi( i\eta + (x_j-x_k))]}{\sin[\pi( i\eta -(x_j-x_k))]}
\label{log Bethe equation q-boson}
\end{eqnarray}
where $I_j$ is (half-)integers when $M$ is (even) odd.
From this formula, we can show the existence and uniqueness of the solutions of the Bethe Ansatz equations once we assign $\{I_1,\cdots,I_M\}$
in the similar manner with the Bose gas model \cite{Yang:1968rm, Korepin:book}.
In \cite{Korff:2013rsa}, Korff proved the completeness of the Bethe vectors in the $q$-boson model with an indeterminate  $t=q^2$.

Finally,  let us consider the inner product between 
$\prod_{a=1}^M B(\mu_a)|0\rangle$
and
$\langle 0|\prod_{a=1}^M C(\nu_a)$:
\begin{eqnarray}
\langle0|\prod_{a=1}^M C(\mu_a)\prod_{a=1}^M B(\nu_a)|0\rangle
\label{inner product}
\end{eqnarray}
where $\{\mu_1,\cdots,\mu_M\}$ and $\{\nu_1,\cdots,\nu_M\}$ are generic complex numbers.
In particular, we give a determinant formula for  the inner product   when either of $\{\mu_1,\cdots,\mu_M\}$ or $\{\nu_1,\cdots,\nu_M\}$ satisfy the Bethe Ansatz equations (\ref{Bethe equation}).
For the methods to derive the determinants of  an inner product,  for example, see \cite{Slavnov:1989, Korepin:book}.
In this paper, we follow Slavnov's derivation \cite{Slavnov:2007}  of the inner product based on the commutation relations of the Yang-Baxter algebra, (\ref{commutationAA}) - (\ref{commutationCB2}).
An advantage of this method is to be able to  apply to a wide class of models.

Let us summarize the results of the inner product for the $q$-boson model from here.
We define  the Bethe vectors which are the eigenvector and its dual eigenvector of the transfer matrix  as follows:
\begin{eqnarray}
|\psi(\{\lambda\}_M)\rangle = \prod_{a=1}^M B(\lambda_a)|0\rangle
\quad\mathrm{and}\quad
\langle \psi(\{\lambda\}_M)| = \langle 0|\prod_{a=1}^M C(\lambda_a)
\label{Bethe state}
\end{eqnarray}
where $\{\lambda_1,\cdots,\lambda_M\}$ satisfies the Bethe  Ansatz equations (\ref{Bethe equation}).
The inner product between the Bethe vector and the generic vector with generic complex spectral parameters $\{\mu_1\cdots\mu_M\}$ is expressed by the determinant formula:
\begin{eqnarray}
\langle\psi(\{\lambda\})|\prod_{a=1}^M B(\mu_a)|0\rangle
&=&\prod_{a=1}^M\Bigl(\frac{\mu_a}{\lambda_a}\Bigl)\cdot
\langle 0|\prod_{a=1}^M C(\mu_a)|\psi(\{\lambda\})\rangle \nonumber\\
&=& \prod_{a=1}^M d(\lambda_a)\cdot\chi^{-1}_M(\{\mu\},\{\lambda\})\cdot
\det_M\Bigl(\frac{\partial}{\partial\lambda_j}\Lambda(\mu_k,\{\lambda\})\Bigl)
\label{scalar product:q-boson}
\end{eqnarray}
where $\Lambda(\mu_k,\{\lambda\})$ is the eigenvalue of the transfer matrix (\ref{transfer matrix eigenvalue}) and $\chi_M(\{\mu\},\{\lambda\})$ is the Cauchy determinant:
\begin{eqnarray}
\chi_M(\{\mu\},\{\lambda\}) =
 \frac{\prod_{a>b}^M (\lambda_a - \lambda_b)(\mu_b - \mu_a)}{\prod_{a,b=1}^M(\mu_a - \lambda_b)}.
\end{eqnarray}
When $\{\mu_1,\cdots,\mu_M\}$ in (\ref{scalar product:q-boson}) moreover satisfies the Bethe Ansatz equations (\ref{Bethe equation}),  we obtain
\begin{eqnarray}
\langle \psi(\{\lambda\}_M)|\psi(\{\lambda\}_M)\rangle
&=& \langle0|\prod_{a=1}^M C(\lambda_a)\prod_{a=1}^M B(\lambda_a)|0\rangle\nonumber\\
&=& \frac{\prod_{a,b=1}^M (\lambda_a t - \lambda_b)}{\prod_{\substack{a,b=1\\a\neq b}}^M (\lambda_a - \lambda_b)}
\cdot\det_M\Phi_{j,k}^{\prime}(\{\lambda\}_M)
\label{q-boson Bethe norm}
\end{eqnarray}
where the Gaudin matrix $\Phi^{\prime}_{j,k} (\{\lambda\}_M) $ is
\begin{eqnarray}
\Phi^{\prime}_{j,k} (\{\lambda\}_M) 
&=&\frac{\partial}{\partial\lambda_k}
\log\Bigl\{\lambda_j^{-L} \cdot
\prod_{\substack{b=1\\b\neq j}}^M\frac{\lambda_j t - \lambda_b}{\lambda_j - \lambda_b t}\Bigl\}\nonumber\\
&=&\delta_{j,k} \Bigl\{-\frac{L}{\lambda_j} 
+ \sum_{b=1}^M\frac{(t^2 - 1)\lambda_b}{(\lambda_jt-\lambda_b)(\lambda_b t -\lambda_j)}\Bigl\}
- \frac{(t^2 - 1)\lambda_j}{(\lambda_j t-\lambda_k)(\lambda_k t -\lambda_j)}.
\label{q-boson Gaudin}
\end{eqnarray}
This norm will become one of the most important quantities when we study the Gauge/Bethe correspondence between the $q$-boson model and the topological field theory.
All the result obtained here for the $q$-boson model reproduce that for the phase model \cite{Okuda:2012nx} in the limit $t\rightarrow 0$.

We comment on  relations between the $q$-boson model and the infinite spin XXZ model.
The Bethe Ansatz equations (\ref{exp Bethe equation}) and the norms (\ref{q-boson Bethe norm}) agree with ones for the infinite spin XXZ model under  the appropriate rescaling of  parameters in the both models when the number of sites is even.
See the algebraic Bethe Ansatz and the inner product for the higher spin XXZ model, e.g. \cite{Deguchi:2009zz, Deguchi:2010zz}.
The agreement of the Bethe Ansatz equations and the norms in the $q$-boson model and in the infinite spin XXZ model may  not be accidental.
This is because  the $q$-oscillator representation is equivalent to the infinite spin limit of spin-$s$ representation in the quantum group. 
In the case of $\mathfrak{su}_q(2)$, this fact is proved in \cite{Kulish:1991bk}.
However, equivalence of the Hamiltonian in the both models is not proved yet.


\section{$G/G$ gauged Wess-Zumino-Witten-matter model}
\label{sec:GWZWM}
In this section, we study a generalization of the Gauge/Bethe correspondence between the $G/G$ gauged WZW model and the phase model discovered by
\cite{Okuda:2012nx}.
In the previous section, we have stated that the phase model is realized as the $t=0$ limit of the $q$-boson model. 
Since the Gauge/Bethe correspondence is a correspondence between  topological gauge theories and quantum integrable systems,
there should exist a topological gauge theory corresponding to the $q$-boson model.
We will show that this topological gauge theory is the $G/G$ gauged WZW model coupled to additional matters. 
From here, we call this model as the $G/G$ gauged Wess-Zumino-Witten-matter model.
The purpose of this section is to investigate various relations between the $G/G$ gauged WZW-matter model and  the $q$-boson model by utilizing the cohomological localization method in a similar way with \cite{Moore:1997dj, Gerasimov:2006zt, Okuda:2012nx, Blau:1993tv, Gerasimov:1993ws, Miyake:2011yr}.

This section is organized as follows.
In section \ref{subsec:GWZWM}, we introduce the $G/G$ gauged WZW-matter model on a genus-$h$ Riemann surface.
Then, we apply the cohomological localization method to the model in order to evaluate the partition function in section \ref{subsec:GWZWM localization}.
Furthermore, we evaluate numerically the partition function in section \ref{subsec:GWZWM numerical}.
In section \ref{subsec:GWZWM Gauge/Bethe}, we establish a correspondence between the partition function of the $G/G$ gauged WZW-matter model and $q$-boson model.
In section \ref{subsec:GWZWM Axiom}, we investigate the mathematical structures from the viewpoint of the Atiyah-Segal axiomatic system \cite{Atiyah:1989vu,Segal:2002ei} and give a relation with a TQFT constructed by Korff \cite{Korff:2013rsa}.
Finally, we generalize the Gauge/Bethe correspondence of the partition function to that of the correlation functions in section \ref{sec:Correlation}.


\subsection{$G/G$ gauged Wess-Zumino-Witten-matter model}
\label{subsec:GWZWM}
In this subsection,  we introduce the $G/G$ gauged WZW-matter model on a genus-$h$ Riemann surface.
Since this model is defined as the $G/G$ gauged WZW model coupled to matters on the Riemann surface,
let us first define the $G/G$ gauged WZW model on a genus $h$ Riemann surface $\Sigma_h$.
See \cite{Blau:1993tv, Witten:1991mm} for the $G/G$ gauged WZW model in details.

The $G/G$ gauged WZW model consists  of a following fields:
a $G$-valued field $g(z,\bar{z})$, a connection  $A = A^{(1,0)}+ A^{(0,1)}$  on a $G$-bundle $E$
and a Grassmann odd one-form  $\lambda = \lambda^{(1,0)} + \lambda^{(0,1)} \in \Omega^{1}\left(\Sigma_h, \text{End}(E)\right)$ 
\footnote{Once we define a complex structure on the Riemann surface,  
the one-form $\lambda$ is decomposed into the (1,0)-form $\lambda^{(1,0)} \in \Omega^{(1,0)}\left(\Sigma_h, \text{End}(E)\right)$ 
and the (0,1)-form $\lambda^{(0,1)} \in \Omega^{(0,1)}\left(\Sigma_h, \text{End}(E)\right)$.}
.
 The action is defined as
\begin{eqnarray}
 && \hspace{-1cm} S_{{\rm GWZW}}(g,A,\lam)\nonumber\\
&=& -\frac{1}{8\pi} \int_{\Sigma_h}  \Tr \left(g^{-1} d_A g\wedge \ast g^{-1}d_A g\right)
 - i\Gamma(g,A) 
 + \frac{i}{4\pi}\int_{\Sigma_h} \Tr(\lambda\wedge\lambda).
\label{GWZW action}
\end{eqnarray}
where $d_A$ is the covariant derivative, $d_A = d g + [A,g]$.
Here, $\Gamma(g,A)$ is the gauge invariant extension of the Wess-Zumino term: 
\begin{eqnarray}
\Gamma(g,A) = \Gamma(g) -\frac{1}{4\pi} \int_{\Sigma_h}   \Tr \left\{A\wedge(d g g^{-1} + g^{-1} d g) + A g^{-1}\wedge A g\right\}
\end{eqnarray}
where the Wess-Zumino term $\Gamma(g)$ is
\begin{eqnarray}
\Gamma(g) = \frac{1}{12\pi} \int_B  \Tr\left( g^{-1} d g \wedge g^{-1} d g  \wedge g^{-1} d g \right).
\end{eqnarray}
Here, $B$ is a certain three dimensional manifold with the Riemann surface at the boundary, $\partial B=\Sigma_h$.

From now on, let us construct the action of the $G/G$ gauged WZW-matter model on a genus-$h$ Riemann surface.
The additional matters are as follows:
$\Phi$ ($\psi$) is a Grassmann even (odd) section of the bundle $\mathrm{End}(E)$, respectively.  
The  auxiliary fields  $\varphi^{(1,0)} \in   \Omega^{(1,0)} (\Sigma_h, \text{End}(E))$ and  $ \varphi^{(0,1)}  \in \Omega^{(0,1)} (\Sigma_h, \text{End}(E))$ are 
Grassmann even.   The  auxiliary fields $\chi^{(1,0)} \in  \Omega^{(1,0)} (\Sigma_h, \text{End}(E))$  and $\chi^{(0,1)} \in \Omega^{(0,1)} (\Sigma_h, \text{End}(E))$ are Grassmann odd
\footnote{Note that  the spin of the matters in this model is different from that in \cite{Gerasimov:2006zt} because of $\Phi, \psi  \in   \Omega^{1} (\Sigma_h, \text{End}(E))$ etc.}
. 

Since the $G/G$ gauged WZW-matter model is a topological field theory, the action of the matter part should be expressed as a BRST-exact term.
The BRST transformation generated by a BRST charge $Q_{(g,t)}$ is defined as
\begin{eqnarray}
&&Q_{(g,t)} A = \lambda, \quad 
Q_{(g,t)} \lambda^{(1,0)} = (A^g)^{(1,0)}- A^{(1,0)},\quad 
Q_{(g,t)} \lambda^{(0,1)} = - (A^{g^{-1}})^{(0,1)}+  A^{(0,1)},\nonumber\\
&&Q_{(g,t)} g=0,\quad
Q_{(g,t)} \Phi = \psi,\quad
Q_{(g,t)} \Phi^{\dagger} = \psi^{\dagger},\quad
Q_{(g,t)} \psi = t g^{-1}\Phi g - \Phi, \nonumber\\ 
&&Q_{(g,t)} \psi^{\dagger} = - t g \Phi^{\dagger}g^{-1} +   \Phi^{\dagger},\quad
Q_{(g,t)} \chi^{(1,0)} = \varphi^{(1,0)},\quad
Q_{(g,t)} \chi^{(0,1)} = \varphi^{(0,1)},\nonumber\\ 
&&Q_{(g,t)} \varphi^{(1,0)} = t g^{-1}\chi^{(1,0)} g - \chi^{(1,0)},\quad
Q_{(g,t)} \varphi^{(0,1)} = - t g \chi^{(0,1)} g^{-1} +  \chi^{(0,1)}
\label{BRST}
\end{eqnarray}
where $0\le t <1$ and $A^g = g^{-1}Ag + g^{-1}dg$.
This is a natural generalization of the BRST transformation in the $G/G$ gauged WZW model.

Moreover,
the square of the BRST transformation $Q_{(g,t)}$ generates the finite gauge and U(1) transformation ${\cal L}_{(g,t)}$, $Q_{(g,t)}^2 = {\cal L}_{(g,t)}$:
\begin{eqnarray}
&&{\cal L}_{(g,t)} A^{(1,0)} = (A^g)^{(1,0)} - A ^{(1,0)},
\quad {\cal L}_{(g,t)} A^{(0,1)} =  -(A^{g^{-1}})^{(0,1)} + A ^{(0,1)},\nonumber\\
&&{\cal L}_{(g,t)} \lambda^{(1,0)} = g^{-1}\lambda^{(1,0)}g - \lambda^{(1,0)},
\quad {\cal L}_{(g,t)} \lambda^{(0,1)} = -g\lambda^{(0,1)}g^{-1} + \lambda^{(0,1)},\nonumber\\
&&{\cal L}_{(g,t)} g =0,\quad
 {\cal L}_{(g,t)} \Phi = t g^{-1}\Phi g - \Phi,\quad
{\cal L}_{(g,t)} \Phi^{\dagger} = - t g \Phi^{\dagger} g^{-1} +  \Phi^{\dagger},\nonumber\\
&& {\cal L}_{(g,t)} \psi = t g^{-1} \psi g - \psi,\quad
{\cal L}_{(g,t)} \psi^{\dagger} = - t g \psi^{\dagger} g^{-1} +  \psi^{\dagger},\nonumber\\
&&{\cal L}_{(g,t)} \chi^{(1,0)} = t g^{-1} \chi^{(1,0)} g - \chi^{(1,0)},\quad
{\cal L}_{(g,t)} \chi^{(0,1)} = - t g \chi^{(0,1)} g^{-1} +  \chi^{(0,1)},\nonumber\\
&&{\cal L}_{(g,t)}\varphi^{(1,0)} = t g^{-1}\varphi^{(1,0)} g -\varphi^{(1,0)},\quad
{\cal L}_{(g,t)}\varphi^{(0,1)} = - t g \varphi^{(0,1)} g^{-1} +  \varphi^{(0,1)}.
\end{eqnarray}

We define the partition function of the $G/G$ gauged WZW-matter model with the level $k$ on $\Sigma_h$ by
\begin{eqnarray}
&& Z^{G}_{\rm GWZWM}(\Sigma_h,k, t)\nonumber\\
&& \quad = \int \D g  \D^2 A \D^2 \lambda \D\Phi \D\Phi^{\dagger} \D \psi \D\psi^{\dagger} \D^2 \varphi \D^2 \chi
e^{-k S_{\rm GWZWM}(g, A, \lambda, \Phi, \Phi^{\dagger}, \psi, \psi^{\dagger}, \varphi, \chi) }
\end{eqnarray}
where the action is defined as
\begin{eqnarray}
&& \hspace{-1cm} S_{\rm GWZWM}(g, A, \lambda, \Phi, \Phi^{\dagger}, \psi, \psi^{\dagger}, \varphi, \chi) \nonumber\\
&& \hspace{1cm} = S_{\rm GWZW}(g, A, \lambda)
 + S_{\rm matter}(g, A,  \Phi, \Phi^{\dagger}, \psi, \psi^{\dagger}, \varphi, \chi).
\label{BRST exact action}
\end{eqnarray}
Here, the matter part of (\ref{BRST exact action}) is represented as the BRST-exact form:
\begin{eqnarray}
S_{\rm matter}(g, A,  \Phi, \Phi^{\dagger}, \psi, \psi^{\dagger}, \varphi, \chi) = Q_{(g,t)}\cdot {\cal R}
\label{matter action}
\end{eqnarray}
with
\begin{eqnarray}
{\cal R} 
= \frac{1}{4\pi}\int_{\Sigma_h} \left\{d\mu\Tr(\Phi^{\dagger}\psi - \Phi\psi^{\dagger} ) 
+ {\cal R}_1 + {\cal R}_2\right\}
\label{R}
\end{eqnarray}
where $d\mu$ is a volume form on the Riemann surface.
${\cal R}_1$ and ${\cal R}_2$  are defined as
\begin{eqnarray}
{\cal R}_1 &=& \Tr\left\{\chi^{(0,1)}\wedge \left( \partial_A\Phi - \Phi X + X \Phi\right)\right\},\\
{\cal R}_2 &=& \Tr\left\{\chi^{(1,0)}\wedge \left({\bar{\partial}}_A \Phi^{\dagger} - Y \Phi^{\dagger}  +  \Phi^{\dagger} Y\right)\right\}
\end{eqnarray}
with
\begin{eqnarray}
X &=& \sum_{n=0}^{\infty} X_n = \sum_{n=0}^{\infty} g^{-n}(g^{-1} \partial_A g)g^n,\\
Y &=& \sum_{n=0}^{\infty} Y_n = \sum_{n=0}^{\infty} g^n({\bar{\partial}}_A g\cdot g^{-1})g^{-n}.
\end{eqnarray}
Here we define the covariant derivatives as $\partial_A f = \partial f + [A^{(1,0)},f]$  and  $\bar{\partial}_A f = \bar{\partial} f + [A^{(0,1)},f]$ for a scalar field $f$. 
When we carry out the BRST transformation for the action (\ref{matter action}), we obtain
\begin{eqnarray}
\hspace{1.5cm}&& \hspace{-1.5cm}S_{\rm matter}(g, A,  \Phi, \Phi^{\dagger}, \psi, \psi^{\dagger}, \varphi, \chi) \nonumber\\
  &=& -\frac{1}{2\pi} \int_{\Sigma_h}d\mu \Tr\left(\Phi\Phi^{\dagger} + \psi\psi^{\dagger} - t \Phi^{\dagger}g^{-1}\Phi g\right)\nonumber\\
&& + \frac{1}{4\pi} \int_{\Sigma_h} \Tr \left\{\varphi^{(0,1)}\wedge ( \partial_A \Phi + [X,\Phi])
-\chi^{(0,1)}\wedge ( \partial_A \psi+ [X,\psi])\right.\nonumber\\
&&\left.
+\varphi^{(1,0)}\wedge ( \bar{\partial}_A \Phi^{\dagger}   - [Y,\Phi^{\dagger}])
-\chi^{(1,0)}\wedge ( \bar{\partial}_A \psi^{\dagger}  - [Y,\psi^{\dagger}])\right\}.
\label{GWZWH action component}
\end{eqnarray}

The partition function of the $G/G$ gauged WZW-matter model is a topological invariant because the $G/G$ gauged WZW-matter model is a topological field theory.
Recall that the partition function of the $G/G$ gauged WZW model counts the number of the conformal blocks of the $G$ WZW model \cite{Witten:1991mm, Witten:1988hf}.
Also, the $G/G$ gauged WZW-matter model is  a one-parameter deformation of  the $G/G$ gauged WZW model as it immediately becomes clear. 
Therefore,  we expect that the partition function of the $G/G$ gauged WZW-matter model counts the number of the building blocks of a certain underlying field theory but we do not know what its field theory is.

\subsection{Localization}
\label{subsec:GWZWM localization}
Let us set the gauge group $G$  as  $U(N)$ 
and  evaluate the partition function of the $U(N)/U(N)$ gauged WZW-matter model by using the cohomological localization method.
Since we can not directly evaluate the partition function with the action  (\ref{BRST exact action}),
we consider a more general action given by
\begin{eqnarray}
&&\hspace{-1cm}S_{\rm matter}^{\tau_1,\tau_2}(g, A, \Phi, \Phi^{\dagger}, \psi, \psi^{\dagger}, \varphi, \chi)\nonumber\\
&&\hspace{-1cm}\quad = Q_{(g,t)}\cdot\Bigl[\frac{1}{4\pi}\int_{\Sigma_h} \Bigl\{d\mu\Tr\left(\Phi^{\dagger} \psi  - \Phi\psi^{\dagger}\right) 
+\tau_1 \left({\cal R}_1 + {\cal R}_2\right) -\tau_2 \Tr(\chi\wedge\ast\varphi)\Bigl\}\Bigl]
\label{deformed GWZWM action}
\end{eqnarray}
where we denote $\ast$ as the Hodge dual operator.
Also, $\varphi = \varphi^{(1,0)} + \varphi^{(0,1)}$ and $\chi = \chi^{(1,0)} + \chi^{(0,1)}$.
For $\tau_1 =1,\tau_2=0$, (\ref{deformed GWZWM action}) matches (\ref{matter action}).
From the viewpoint of the cohomological localization for the path integral,
we  expect that the partition function for  $\tau_1=1,\tau_2=0$ coincides with  that for $\tau_1=0,\tau_2=1$.
Thus, we consider the case of $\tau_1=0,\tau_2=1$ from now on.
In this case, the action (\ref{deformed GWZWM action}) becomes 
\begin{eqnarray}
&&\hspace{-1cm} S_{\rm matter}^{\tau_1=0, \tau_2=1}(g, A, \Phi, \Phi^{\dagger}, \psi, \psi^{\dagger}, \varphi, \chi)\nonumber\\
&=&Q_{(g,t)}\cdot \left[
\frac{1}{4\pi} \int_{\Sigma_h}  \Bigl\{   d\mu \Tr(\Phi^{\dagger}\psi- \Phi \psi^{\dagger}) -   \Tr(\chi\wedge\ast\varphi) \Bigl\}\right]\nonumber\\
&=&- \frac{1}{2\pi}\int_{\Sigma_h} d\mu
\Tr\Bigl\{\Phi\Phi^{\dagger} - t\Phi g \Phi^{\dagger}g^{-1} 
+\psi\psi^{\dagger}\Bigl\} \nonumber\\
&&- \frac{1}{2\pi}\int_{\Sigma_h}d^2z \Tr\left(\varphi^{(1,0)}\wedge\ast\varphi^{(0,1)} - \chi^{(1,0)}\wedge\ast\chi^{(0,1)} + t\chi^{(1,0)} g \wedge\ast \chi^{(0,1)} g^{-1}\right).
\label{deformed GWZWM action3}
\end{eqnarray}
The action is going to become quadratic in terms of $\Phi$, $\varphi$, $\psi$ and $\psi$ after we take a diagonal gauge.
Therefore, we can evaluate the partition function in a similar manner with \cite{Blau:1993tv}.
For simplicity of notation, we denote this action as $S_{\rm matter}(g,\Phi, \Phi^{\dagger}, \psi, \psi^{\dagger}, \varphi, \chi)$ from here.

As stated in the previous subsection, the $G/G$ gauged WZW-matter model is a one-parameter deformation of the $G/G$ gauged WZW model.
Here, let us explain this.
In the action (\ref{deformed GWZWM action3}), we find  that the interaction terms between the fields in the $G/G$ gauged WZW model and the additional matters disappear when we set $t=0$. 
Hence, the $G/G$ gauged WZW-matter model becomes the $G/G$ gauged WZW model by integrating out the matter part at $t=0$.
Therefore, we can regard the $G/G$ gauged WZW-matter model  as  a one-parameter deformation of the $G/G$ gauged WZW model.

Let us take a diagonal gauge $g(z,\bar{z}) \equiv \exp\left\{2\pi i\sum_{a=1}^N \phi_a(z,\bar{z}) H^a\right\}$ where $H^1, \cdots, H^N$ are the  Cartan generators of $U(N)$
and $0 \le \phi_1, \cdots, \phi_N < 1$.
Then, the partition function under the diagonal gauge becomes
\begin{eqnarray}
&&Z_{\rm GWZWM}^{U(N)} (\Sigma_h, k, t)\nonumber\\ 
&&\quad = \frac{1}{|W|}\int \D^2 A \D^2 \lambda \D\phi \D\Phi \D\Phi^{\dagger} \D\psi \D\psi^{\dagger} \D^2 \varphi \D^2\chi \Det(1- \Ad(e^{2\pi i \phi}))\nonumber\\
&&\quad\quad \times\exp\left\{-k S_{\rm GWZW}(\phi,A)
-k S_{\rm matter}(\phi,\Phi, \Phi^{\dagger}, \psi, \psi^{\dagger}, \chi,\varphi)\right\}
\label{Gauged WZWH partition function}
\end{eqnarray}
where $|W|$ is the order of the Weyl group of $U(N)$ and $\Det(1- \Ad(e^{2\pi i \phi}))$
 is the Faddeev-Popov determinant for the diagonal gauge fixing.
$\Det$ represents the functional determinant.
See \cite{Okuda:2012nx, Blau:1993tv}.

From now on, we explicitly carry out the path integration of (\ref{Gauged WZWH partition function}).
First, we consider the path integral with respect to the connection $A$ and $\lambda$:
\begin{eqnarray}
\int \D^2 A \D^2 \lambda \Det(1- \Ad(e^{2\pi i \phi})) \exp{\left(-k S_{\rm GWZW}(\phi, A, \lambda)\right)}.
\end{eqnarray}
We already evaluated this path integration at \cite{Okuda:2012nx, Blau:1993tv, Gerasimov:1993ws}.
The resulting expression is given by
\begin{eqnarray}
&&\int \prod_{a=1}^N \D^2 A_a\prod_{a=1}^N \D^2 \lambda_a \prod_{\substack{a,b=1\\a \neq b}}^N \bigl(1- e^{2\pi i (\phi_a- \phi_b)}\bigr)^{1-h}
\exp\left\{\sum_{a=1}^{N} \frac{i}{4\pi} \int_{\Sigma_h}\lambda_a\wedge\lambda_a\right\}\nonumber\\
&&\quad\times
\exp\left\{ i \sum_{a=1}^N \int_{\Sigma_h} F_a
    \left((N+k)\phi_a - \sum_{b=1}^N\phi_b + \frac{N-1}{2}
    \right)
\right\}.
\label{GWZW path-integral}
\end{eqnarray}
Here, we have expanded an adjoint field $f$ by the Cartan-Weyl basis as
\begin{eqnarray}
f &=&\sum_{a=1}^N f_a (i H^a) + \sum_{\alpha \in \Delta} f_{\alpha}  (i E^{\alpha})
\end{eqnarray}
where $\alpha$ is a root and $\Delta$ represents the set of all roots.

Next, we evaluate the path integration with respect to $\Phi$, $\Phi^{\dagger}$, $\psi$, $\psi^{\dagger}$, $\varphi$ and $\chi$:
\begin{eqnarray}
\int  \D\Phi \D\Phi^{\dagger} \D\psi \D\psi^{\dagger} \D^2 \varphi \D^2 \chi 
\exp\left(-k S_{\rm matter}(\phi,\Phi, \Phi^{\dagger}, \psi, \psi^{\dagger}, \chi,\varphi)\right).
\end{eqnarray}
The action of the matter part under the diagonal gauge is expressed as
\begin{eqnarray}
&&\hspace{-1cm} S_{\rm matter}(\phi, \Phi, \Phi^{\dagger}, \psi, \psi^{\dagger}, \varphi, \chi)\nonumber\\
&=&  -\frac{1}{2\pi}\int_{\Sigma_h}d\mu \Tr (\psi\psi^{\dagger})
-  \frac{1}{2\pi}\int_{\Sigma_h}\Tr (\varphi^{(1,0)}\wedge\ast\varphi^{(0,1)})+\nonumber\\
&& +\frac{1}{2\pi}\int_{\Sigma_h} d\mu 
\Bigl\{
 (1-t)\sum_{a=1}^N \Phi_a\Phi^{\dagger}_a 
+ \sum_{\alpha\in \Delta}\left(1-t e^{2\pi i\alpha(\phi)}\right) \Phi_{-\alpha}\Phi^{\dagger}_{\alpha}\Bigl\}\nonumber\\
&&-\frac{1}{2\pi} \int_{\Sigma_h} \Bigl\{ (1-t) \sum_{a=1}^N  \chi^{(1,0)}_a \wedge \ast \chi_a^{(0,1)}
+ \sum_{\alpha\in \Delta} (1-t e^{2\pi i\alpha(\phi)})  \chi^{(1,0)}_{-\alpha}\wedge\ast\chi^{(0,1)}_{\alpha}\Bigl\}
\label{deformed GWZWM action4}
\end{eqnarray}
where $\alpha(\phi) = \sum_{a=1}^N\alpha_a \phi_a$.
By performing the path integral with respect to $\chi_{\alpha}^{(1,0)}$ and  $\chi_{\alpha}^{(0,1)}$,  we obtain
\begin{eqnarray}
&&\int  \prod_{\alpha \in \Delta} \D \chi_{\alpha}^{(1,0)} \prod_{\alpha \in \Delta} \D \chi_{\alpha}^{(0,1)} \prod_{\alpha \in \Delta} 
\exp\left\{-\frac{k}{2\pi}\int_{\Sigma_h}  \chi^{(0,1)}_{\alpha} (1-t e^{2\pi i\alpha(\phi)}) \wedge\ast \chi^{(1,0)}_{-\alpha}\right\}\nonumber\\
&&=\int \prod_{\alpha>0} \D \chi^{(1,0)}_{\alpha} \prod_{\alpha>0} \D \chi^{(1,0)}_{-\alpha} \prod_{\alpha>0} \D \chi^{(0,1)}_{\alpha} \prod_{\alpha>0} \D \chi^{(0,1)}_{-\alpha}  \nonumber\\
&&\hspace{1cm}\times\prod_{\alpha>0} \exp\left\{-\frac{k}{2\pi}\int _{\Sigma_h} \left( \chi_{\alpha}^{(0,1)} M_{\alpha}(t)  \chi^{(1,0)}_{-\alpha}
+\chi_{-\alpha}^{(0,1)} M_{-\alpha}(t)  \chi^{(1,0)}_{\alpha}\right)\right\}\nonumber\\
&& = \prod_{\alpha>0}\Det_{(1,0)} M_{\alpha}(t)\cdot \prod_{\alpha > 0} \Det_{(1,0)} M_{-\alpha}(t)
\label{chi}
\end{eqnarray}
where $M_{\alpha}(t) = 1 - t e^{2\pi i\alpha(\phi)}$.
Furthermore,   by performing the path integral with respect to $\Phi_{\alpha}$ and $\Phi^{\dagger}_{\alpha}$,  we obtain
\begin{eqnarray}
&& \int \prod_{\alpha \in \Delta} \D \Phi_{\alpha} \prod_{\alpha\in \Delta}  \D \Phi_{\alpha}^{\dagger} \prod_{\alpha\in \Delta}
 \exp\left\{-\frac{k}{2\pi}\int d\mu\Phi^{\dagger}_{\alpha} \left(1-t e^{2\pi i\alpha(\phi)}\right) \Phi_{-\alpha}\right\}\nonumber\\
 &&=  \int \prod_{\alpha>0}\D \Phi_{\alpha} \prod_{\alpha>0}\D \Phi_{-\alpha}\prod_{\alpha>0}\D \Phi_{\alpha}^{\dagger}\prod_{\alpha>0}\D \Phi_{-\alpha}^{\dagger} \nonumber\\
&&\hspace{1cm}\times \prod_{\alpha>0}\exp\left\{-\frac{k}{2\pi}\int d\mu
 \left(\Phi^{\dagger}_{\alpha} M_{\alpha}(t) \Phi_{-\alpha} +\Phi^{\dagger}_{-\alpha} M_{-\alpha}(t) \Phi_{\alpha}\right) \right\}\nonumber\\
 &&= \prod_{\alpha>0}\left[\Det_0 M_{\alpha}(t)\right]^{-1}\cdot \prod_{\alpha > 0} \left[\Det_0 M_{-\alpha}(t)\right]^{-1}.
\label{Phi}
\end{eqnarray}
Putting together with (\ref{chi}) and  (\ref{Phi}), the contributions to the partition function from $\Phi_{\alpha}$, $\Phi_{\alpha}^{\dagger}$, $\chi^{(1,0)}_{\alpha}$ and $\chi_{\alpha}^{(0,1)}$ become
\begin{eqnarray}
\prod_{\alpha>0} \frac{\Det_{(1,0)} M_{\alpha}(t)}{\Det_0 M_{\alpha}(t)}
\times
\prod_{\alpha>0}\frac{\Det_{(1,0)} M_{-\alpha}(t)}{\Det_0 M_{-\alpha}(t)}.
\label{ratio}
\end{eqnarray}
Recall that our gauge fixing is partial and the abelian gauge symmetry remains as the residual symmetry.
Therefore, we evaluate this ratio of the functional determinant by using the heat kernel regularization, which respects the abelian gauge symmetry, 
for the twisted Dolbeault complex as well as the case of the gauged WZW model.
The difference of the regularized traces  is evaluated as follows \cite{Blau:1993tv}:
\begin{eqnarray}
 &&\lim_{T \to 0}\left\{ \mathrm{Tr}_{ \Omega^{0} ( \Sigma_h, \mathrm{End}(E)_{\alpha})}  \left( e^{-T \Delta} \log M_{\alpha}(t) \right) 
-\mathrm{Tr}_{ \Omega^{(1,0)} (\Sigma_h, \mathrm{End}(E)_{\alpha}) } \left( e^{-T \Delta} \log M_{\alpha}(t)  \right)\right\} \nonumber \\
&& \hspace{0.5cm} = \left\{\frac{1}{8\pi} \int_{\Sigma_h} R 
+ \frac{1}{2\pi}\int_{\Sigma_h}\alpha_{\ell}F^{\ell}\right\} \log{M_{\alpha}(t)}
\end{eqnarray}
where $R$ denotes a scalar curvature on a genus-$h$ Riemann surface.
Here,  $\text{End}(E)_{\alpha}$ is the restriction of  $\text{End}(E)$  into $E^{\alpha}$, and  $\Delta $ is the Laplace operator with the coefficient $\text{End}(E)_{\alpha}$.
Then, (\ref{ratio}) is evaluated as
\begin{eqnarray}
\prod_{\alpha>0}
\exp\left\{- \frac{1}{8\pi}\int_{\Sigma_h}R\log M_{\alpha}(t)M_{-\alpha}(t) 
-  \frac{1}{2\pi}\int_{\Sigma_h}\alpha_{\ell}F^{\ell}\log\frac{M_{\alpha}(t)}{M_{-\alpha}(t)}\right\}.
\label{chi-psi}
\end{eqnarray}
By performing the path integration in terms of $\Phi_a$, $\Phi^{\dagger}_a$, $\chi_a^{(1,0)}$ and $\chi_a^{(0,1)}$,
we furthermore obtain
\begin{eqnarray}
\prod_{a=1}^N(1-t)^{h-1}.
\label{chi-psi weight}
\end{eqnarray}
Also, the contribution to the partition function from $\varphi$ and $\psi$ cancel out.

Together with (\ref{GWZW path-integral}), (\ref{chi-psi}) and (\ref{chi-psi weight}), the resulting expression for the partition function (\ref{Gauged WZWH partition function}) is
\begin{eqnarray}
&&\hspace{-1cm} Z^{U(N)}_{\text{GWZWM}}(\Sigma_h,k,t)\nonumber\\
\hspace{1cm}&&=\frac{1}{|W|} \int \prod_{a=1}^N \D \phi_a \prod_{a=1}^N \D^2 \lambda_a \prod_{a=1}^N \D^2 A_a
 \left( \frac{\prod_{\substack{a,b=1\\a \neq b}}^N \bigl(1- e^{2\pi i (\phi_a- \phi_b)}\bigr)}
 {\prod_{a,b=1}^N \bigl(1- t e^{ 2\pi i (\phi_a- \phi_b)}\bigr)}\right)^{1-h}\nonumber\\
\hspace{1cm}&&\times  \exp  \left\{  i \sum_{a=1}^{N}   
\int_{\Sigma_h} \left( \beta_a(\phi) F_a + \frac{k}{4\pi} \lambda_a \wedge \lambda_a\right) \right\}\label{incorrect partition function}
\end{eqnarray}
where $\beta_a (\phi)$ is defined by
\begin{eqnarray}
\beta_a (\phi) =  k \phi_a - \frac{i}{2\pi} \sum_{\substack{b=1\\b\neq a}}^N \log \left( \frac{e^{2\pi i\phi_a}- t e^{2\pi i \phi_b}}{t e^{2\pi i\phi_a}- e^{2\pi i \phi_b } } \right).
\end{eqnarray}
Here, we  have used the fact that the constant modes of $\{\phi_1,\cdots,\phi_N\}$ only contribute to the partition function as we will show later,
and therefore $\frac{1}{8\pi}\int_{\Sigma_h} R=1-h$
\footnote{Since the term with the scalar curvature $R$  does not break the abelianized BRST invariance, we can simply replace $\phi_1, \cdots, \phi_N$ by constant.} .

Let us define an abelianized effective action by
\footnote{We do not include into the abelianized effective action
\begin{eqnarray}
 \left( \frac{\prod_{\substack{a,b=1\\a \neq b}}^N \bigl(1- e^{2\pi i (\phi_a- \phi_b)}\bigr)}
 {\prod_{a,b=1}^N \bigl(1- t e^{ 2\pi i (\phi_a- \phi_b)}\bigr)}\right)^{1-h},
 \label{1-loop CS}
\end{eqnarray}
because this term does not break the abelianized BRST invariance and does not affect following results.}
\begin{eqnarray}
S_{\rm eff}(\phi,A,\lambda)
=-i \sum_{a=1}^{N}   
\int_{\Sigma_h} \left( \beta_a(\phi) F_a +\frac{k}{4\pi} \lambda_a \wedge \lambda_a\right).
\label{effective action GWZWM}
\end{eqnarray}
Then, we find that this is not invariant under the abelianized BRST transformation of (\ref{BRST}):
\begin{eqnarray}
Q A_{a}=\lambda_a,\quad
Q \lambda_a =2\pi d \phi_a,\quad
Q \phi_a=0
\label{abelian BRST GWZWM}
\end{eqnarray}
where $Q$ is the abelianized BRST charge.
A reason why the effective action (\ref{effective action GWZWM}) is not invariant under the  BRST transformation, is considered as follows.
 The heat kernel regularization scheme respects  the abelianized gauge symmetry but not the  BRST symmetry,  and therefore breaks it. 
Since the regularization scheme breaks the BRST symmetry, we have to add  counterterms to restore the BRST symmetry.
A prescription to restore the BRST symmetry is given by \cite{Gerasimov:2006zt}. That is to modify the effective action such that it satisfies decent equations
\footnote{In  \cite{Miyake:2011yr}, the volume of vortex moduli space is evaluated by using the cohomological localization method. When the gauge group is U(1),  the volume calculated by the localization can not reproduce one obtained in \cite{Manton:1998kq} unless one modifies the effective action by following the prescription. Thus, in our model, we consider that it is necessary to restore the BRST invariance in the effective action by following the prescription.}.

Now, we explain the decent equations and how to restore the BRST invariance of the effective action. 
First, we define a local operator $\mathcal{O}^{(0)}$ as
\begin{eqnarray}
\mathcal{O}^{(0)}=W(\phi)
\end{eqnarray}
where $W(\phi)$ is an arbitrary function of  $\phi_1,\cdots,\phi_N$ on the Riemann surface.
Here, we introduce the descend equations:
\begin{eqnarray}
d \mathcal{O}^{(n-1)} = Q \mathcal{O}^{(n)}.
\label{descent}
\end{eqnarray}
where $\mathcal{O}^{(n)}$, $n=0, 1, 2$, are defined as $n$-form valued local operators.
Note that the 3-form local operator $\mathcal{O}^{(3)}$ does not exist because we consider the Riemann surface as the base manifold.
If ${\cal O}^{(n)}$ satisfies the descend equations, we find that the  integration of  $\mathcal{O}^{(n)}$ over a $n$-cycle $\gamma_n$, namely $\int_{\gamma_n} \mathcal{O}^{(n)}$, becomes the BRST-closed operator under the abelianized BRST transformation (\ref{abelian BRST GWZWM}):
\begin{eqnarray}
Q\cdot\int_{\gamma_n} \mathcal{O}^{(n)} = 0.
\end{eqnarray}
We can concretely construct the BRST-closed operators as follows:
\begin{eqnarray}
{\cal O}^{(0)} &=& W(\phi),\nonumber\\
{\cal O}^{(1)} &=& \frac{1}{2\pi}\sum_{a=1}^N \frac{\partial W(\phi)}{\partial \phi_a}\lambda_a,\nonumber\\
{\cal O}^{(2)} &=& \frac{1}{8\pi^2}\sum_{a,b=1}^N \frac{\partial^2 W(\phi)}{\partial \phi_a \partial \phi_b}\lambda_a\wedge\lambda_b 
+ \frac{1}{2\pi}\sum_{a=1}^N \frac{\partial W(\phi)}{\partial \phi_a}F_a.
\end{eqnarray}
In our case, by defining  the function $W(\phi)$ as 
\begin{eqnarray}
\frac{1}{2\pi}\frac{\partial W(\phi)}{\partial \phi_a} = \beta_a(\phi),
\end{eqnarray}
we find that the operator $\mathcal{O}^{(2)}$ becomes
\begin{eqnarray}
 \mathcal{O}^{(2)} =\sum_{a=1}^N
 \left(\beta_a(\phi)  F_a + \frac{1}{4\pi} \sum_{b=1}^N \frac{\partial \beta_{b}(\phi)}{\partial \phi_a} \lambda_a \wedge \lambda_b\right).
 \label{true effective action}
\end{eqnarray}
In order to restore the BRST invariance in the effective action (\ref{effective action GWZWM}), 
we must replace (\ref{effective action GWZWM}) with (\ref{true effective action}):
\begin{eqnarray}
S_{\rm eff}(\phi,A,\lambda) = - i \sum_{a=1}^{N} \int_{\Sigma_h}
   \left( \beta_a(\phi)  F_a 
   + \frac{1}{4\pi} \sum_{b=1}^N \frac{\partial \beta_{b}(\phi)}{\partial \phi_a} \lambda_a \wedge \lambda_b\right).
\end{eqnarray}
As a result, we  can restore the BRST symmetry in the effective theory under this replacement and obtain the following expression for the partition function:
\begin{eqnarray}
\hspace{-1cm}&& Z^{U(N)}_{\text{GWZWM}}(\Sigma_h,k,t)\nonumber\\
&&\hspace{1cm} =\frac{1}{|W|}\int \prod_{a=1}^N \D \phi_a \prod_{a=1}^N \D^2 \lambda_a  \prod_{a=1}^N \D^2 A_a
 \left( \frac{\prod_{\substack{a,b=1\\a \neq b}}^N \bigl(1- e^{ 2\pi i (\phi_a- \phi_b)}\bigr)}
 {\prod_{a,b=1}^N \bigl(1- t e^{ 2\pi i (\phi_a- \phi_b)}\bigr)}\right)^{1-h}  \nonumber\\
 &&\hspace{1cm}\times \exp  \left\{  i \sum_{a=1}^{N} \int_{\Sigma_h}
   \left( \beta_a(\phi)  F_a 
   + \frac{1}{4\pi} \sum_{b=1}^N \frac{\partial \beta_{b}(\phi)}{\partial \phi_a} \lambda_a \wedge \lambda_b\right) \right\}.
\end{eqnarray}

Let us see that the field configurations of $\phi_1,\cdots, \phi_N$ reduce to the constant configurations.
In order to see this, a two-form field strength $F_b$  be decomposed to the harmonic part $F^{(0)}_b$ and the exterior derivative of a one-form $da_b$ by the Hodge decomposition theorem,
$F_b = F_b^{(0)} + d a_b$.
Integrating the harmonic part of the field strength gives the $b$-th diagonal $U(1)$-charge $k_b$ of the background gauge fields:
\begin{eqnarray}
\frac{1}{2\pi}\int_{\Sigma_h} F^{(0)}_b = k_b.
\label{magnetic charge2}
\end{eqnarray}
We subsequently decompose the 1-form fermion $\lambda$ into $\lambda_a=\lambda^{(0)}_a + \delta\lambda_a$ in the same way as the field strength.
Here, $\lambda_a^{(0)}$ is a harmonic 1-form fermion and $\delta\lambda_a$ is  fluctuations orthogonal to $\lambda_a^{(0)}$, $\lambda_a^{(0)}\wedge \delta \lambda =0$.
Next, we integrate $a_b$ by parts.
Then, we find that the contribution from the path integration of $\delta\lambda_b$  completely cancel out with a Jacobian for the change of variables of $a_b$.
We subsequently  integrate out $a_b$ and obtain a delta functional of $d\phi_a$.
By performing the path integral with respect to $\phi_a$, we find that the field configurations of $\phi_1,\cdots, \phi_N$ reduce to the constant configuration.

Since the number of fermionic zero-modes of each $\lambda_a^{(0)}$ is equal to the number of the harmonic forms $2h$ on the genus-$h$ Riemann surface,
performing the path integration with respect to $\lambda_1^{(0)},\cdots,\lambda_N^{(0)}$ gives an additional factor
\begin{eqnarray}
\mu_q(\phi)^h = \left|\det\left(\frac{\partial \beta_{b}(\phi)}{\partial \phi_a}\right)\right|^h.
\end{eqnarray}
Therefore, the resulting expression for  the partition function is
\begin{eqnarray}
Z^{U(N)}_{\text{GWZWM}}(\Sigma_h,k,t) &=&\frac{1}{|W|} \sum_{k_1,\cdots,k_N =- \infty}^{\infty} 
\int \prod_{a=1}^N d\phi_a  \mu_q(\phi)^h  e^{i\sum_{a=1}^{N} k_a \beta_a(\phi)} \nonumber\\
&& \hspace{6.5mm}
\times \left(\frac{1}{(1-t)^N}
\prod_{\substack{a,b=1\\a \neq b}}^N  \frac{e^{2\pi i\phi_a} - e^{2\pi i \phi_b}}
 {e^{2\pi i\phi_a} - t e^{2\pi i\phi_b}}\right)^{1-h}.
   \label{partition function aa}
\end{eqnarray}
By using the Poisson resummation formula, we rewrite (\ref{partition function aa}) as
\begin{eqnarray}
Z^{U(N)}_{\text{GWZWM}}(\Sigma_h,k,t)&=&\frac{1}{|W|}\sum_{\ell_1,\cdots,\ell_N = - \infty}^{\infty} 
\int \prod_{a=1}^N d\phi_a \prod_{a=1}^N \delta\left(\beta_a(\phi) - \ell_a\right)
\mu_q(\phi)^h  \nonumber\\
&&\hspace{1cm}\times
\left(\frac{1}{(1-t)^N}
\prod_{\substack{a,b=1\\a \neq b}}^N  \frac{e^{2\pi i\phi_a} - e^{2\pi i \phi_b}}
 {e^{2\pi i\phi_a} - t e^{2\pi i\phi_b}}\right)^{1-h}.
 \label{poisson partition}
\end{eqnarray}
Here, we have utilized a property about the delta function
\begin{eqnarray}
\delta(f(x)) = \sum_i \frac{1}{|f^{\prime}(x_i)|}\delta(x - x_i)
\end{eqnarray}
where $x_i$ is solutions of $f(x)=0$.
Thus, we find that the delta function in the partition function (\ref{poisson partition}) gives an additional factor $\mu_q(\phi)^{-1}$ and constraints for $\phi_1,\cdots,\phi_N$:
\begin{eqnarray}
2\pi i k \phi_a + \sum_{\substack{b=1\\b\neq a}}^N \log \left( \frac{e^{2\pi i\phi_a}- t e^{ 2\pi i \phi_b}}{t e^{2\pi i\phi_a}- e^{2\pi i \phi_b } } \right) 
= 2\pi i \ell_a.
\label{localization configuration}
\end{eqnarray}
If  the solutions of (\ref{localization configuration}) exist in the region $0\le\phi_1,\cdots,\phi_N<1$,  we must sum up   all of them.
In our case,  we can show that the solution is unique up to permutations of $\phi_1,\cdots,\phi_N$.
The partition function is invariant under the permutation and the contributions from it  cancel  out the order of the Weyl group $|W|$.
Therefore, we can generally  set $\phi_1,\cdots,\phi_N$ as  $0\le \phi_1<\cdots<\phi_N<1$.

 By integrating the partition function (\ref{poisson partition}) with respect to $\phi_1,\cdots,\phi_N$, we obtain the final expression for the partition function:
\begin{eqnarray}
\hspace{-0.5cm}Z^{U(N)}_{\text{GWZWM}}(\Sigma_h,k,t) = \sum_{\{\phi_1,\cdots,\phi_N\}\in\{\rm Sol\}}\left\{(1-t)^N \mu_q(\phi) 
 \prod_{\substack{a,b=1\\a \neq b}}^N
 \frac{e^{2\pi i\phi_a} - t e^{2\pi i\phi_b}}
 { e^{2\pi i\phi_a} - e^{2\pi i \phi_b}}\right\}^{h-1}
\end{eqnarray}
where  $\{\rm Sol\}$ represents the set of the solutions which satisfy $0\le \phi_1<\cdots <\phi_N<1$ and the constraint (\ref{localization configuration}).
Also, we can  express explicitly $\mu_q(\phi)$ as
\begin{eqnarray}
\hspace{-1.5cm}\mu_q(\phi)
&=& \det_N \left(\frac{\partial \beta_{b}(\phi)}{\partial \phi_a}\right)\nonumber\\
&=& \det_N\left[\left\{ k - \sum_{c=1}^N
\frac{(t^2-1) e^{2\pi i(\phi_b + \phi_c)}}{(t e^{2\pi i \phi_b} - e^{2\pi i \phi_c})(t e^{2\pi i \phi_c} - e^{2\pi i \phi_b})}
\right\}\delta_{a,b}\right.\nonumber\\
&& \left.\hspace{3cm} +\frac{(t^2-1) e^{2\pi i(\phi_a + \phi_b)}}{(t e^{2\pi i \phi_a} - e^{2\pi i \phi_b})(t e^{2\pi i \phi_b} - e^{2\pi i \phi_a})}\right].
\label{one-loop det}
\end{eqnarray}
Thus, we find that the path integral for  the $U(N)/U(N)$ gauged WZW-matter model reduces to the finite sum of the solutions which satisfy the localized configurations.

Finally, we comment about the normalization of the partition function.
The partition function with the general normalization becomes
\begin{eqnarray}
&&Z^{U(N)}_{\text{GWZWM}}(\Sigma_h,k,t) \nonumber\\
&&\hspace{1cm}= \alpha(t)\beta(t)^{1-h} \sum_{\{\phi_1,\cdots,\phi_N\}\in\{\rm Sol\}}\left\{(1-t)^N \mu_q(x) 
 \prod_{\substack{a,b=1\\a \neq b}}^N
 \frac{e^{2\pi i\phi_a} - t e^{2\pi i\phi_b}}
 { e^{2\pi i\phi_a} - e^{2\pi i \phi_b}}\right\}^{h-1} 
\end{eqnarray}
where $\alpha(t)$ and $\beta(t)$ are arbitrary functions of $t$.
Note that this partition function should coincide with the result in \cite{Okuda:2012nx} in the limit $t\rightarrow 0$ at least.
However, we can not completely determine the normalization of the partition function of the $U(N)/U(N)$ gauged WZW-matter model unlike the gauged WZW model.

\subsection{Numerical simulation}
\label{subsec:GWZWM numerical}
In this subsection, we evaluate numerically the partition function for the $SU(N)/SU(N)$ gauged WZW-matter model with the level $k$ on the genus-$h$ Riemann surface.
Since we  have not determined the normalization of the partition function as discussed  in the previous section,
we assume that the normalization of the partition function of the gauged WZW-matter model coincides with that of the gauged WZW model.
That is to say, we assume that the partition function of the $U(N)/U(N)$ gauged WZW-matter model is
\footnote{In $t=0$, note that the normalization in (\ref{normalized partition U(N)}) is different from that in \cite{Okuda:2012nx}
but the partition function in \cite{Okuda:2012nx} coincides with that in (\ref{normalized partition U(N)}).
This is because  we have interchanged the level $k$ with the dual Coxeter number $N$ in the process of the calculations of the partition function by means of the level-rank duality for the partition function of the $U(N)/U(N)$ gauged WZW model.
\label{footnote}}
\begin{eqnarray}
&&\hspace{-10mm}Z_{\text{GWZWM}}^{U(N)}(\Sigma_h,k,t)\nonumber\\
&&\hspace{-10mm}\quad = \left(\frac{k+N}{k}\right)^h  \sum_{\{\phi_1,\cdots,\phi_N\}\in\{\rm Sol\}}\left\{(1-t)^N \mu_q(x) 
 \prod_{\substack{a,b=1\\a \neq b}}^N
 \frac{e^{2\pi i\phi_a} - t e^{2\pi i\phi_b}}
 { e^{2\pi i\phi_a} - e^{2\pi i \phi_b}}\right\}^{h-1} .
 \label{normalized partition U(N)}
\end{eqnarray}
Also, we assume that the partition function of the $SU(N)/SU(N)$ gauged WZW-matter model
is
\begin{eqnarray}
&&Z_{\text{GWZWM}}^{SU(N)}(\Sigma_h,k,t)\nonumber\\
&& \quad = \left(\frac{N}{k}\right)^h  \sum_{\{\phi_1,\cdots,\phi_N\}\in\{\rm Sol\}}\left\{(1-t)^N \mu_q(x) 
 \prod_{\substack{a,b=1\\a \neq b}}^N
 \frac{e^{2\pi i\phi_a} - t e^{2\pi i\phi_b}}
 { e^{2\pi i\phi_a} - e^{2\pi i \phi_b}}\right\}^{h-1}.
 \label{normalized partition SU(N)}
\end{eqnarray}
As compared (\ref{normalized partition U(N)}) with (\ref{normalized partition SU(N)}), the partition function for the case of  $U(N)$ multiplies that for the case of $SU(N)$ by  $((k+N)/N)^h$.
Thus, we only evaluate numerically the value of the partition function of the $SU(N)/SU(N)$ gauged WZW-matter model  by utilizing e.g. Mathematica.

\begin{table}[t]
\begin{center}
\begingroup
\renewcommand{\arraystretch}{1.3}
  \begin{tabular}{|c|c|c|c|} \hline
  Genus & $k$ & $N$ & Partition Function  \\ \hline\hline
 2 & 2 &  2& $ (1-t)^2 (10 + 10t)$ \\
   & 3 &  & $(1-t)^2 (20 + 16 t)$ \\ 
   & 4 &  & $ (1-t)^2  (35 + 20 t +  t^2)$ \\ 
   & 5 &  & $ (1-t)^2  (56 + 20 t + 4 t^2)$ \\ 
   & 6 &  & $(1-t)^2  (84 + 14 t + 10 t^2)$ \\ 
   & 7 &  & $ (1-t)^2  (120+ 20t^2)$ \\ 
   & 8 &  & $(1-t)^2  (165 - 24 t + 35 t^2)$ \\ 
   & 9 &  & $(1-t)^2  (220 - 60 t + 56 t^2)$ \\ 
   & 10 &  & $(1-t)^2  (286 - 110 t +84t^2)$ \\ \hline
   \end{tabular}
\endgroup
\caption{The partition function  of the $SU(2)/SU(2)$ gauged WZW-matter model with the level $k$ on the genus-$2$ Riemann surface.} 
\label{table:su(2) GWZWM genus 2}
\vspace{1cm}
\begingroup
\renewcommand{\arraystretch}{1.3}
  \begin{tabular}{|c|c|c|c|} \hline
  Genus & $k$ & $N$ & Partition Function \\ \hline\hline
   0 & 2 & 2 & $(1-t)^{-2} (1+t)^{-1}$\\
   1 &    &    & $3$\\
   2 &  &  & $ 10 (1-t)^2 (1 + t)$ \\ 
   3 &  & & $36 (1-t)^4 (1+t)^2$ \\ 
   4 &  &  & $136 (1 - t)^6 (1 + t)^3$ \\ 
   5 & & & $528 (1 - t)^8 (1 + t)^4$ \\ \hline
   \end{tabular}
\endgroup
\caption{The partition function  of the $SU(2)/SU(2)$ gauged WZW-matter model with the level $k=2$ on the genus-$h$ Riemann surface.}
\label{table:su(2) GWZWM higher genus}
   \end{center}
   \end{table}

From now on, we consider the partition function of the $SU(N)/SU(N)$ gauged WZW model with a special level and rank.
First, let us consider the case of genus-$1$,  torus.
In the gauged WZW model, the partition function counts the number of the WZW primary fields and
its number is $(N+k-1)!/(N-1)! k!$.
In the gauged WZW-matter model, we similarly expect that the partition function counts the number of  fields in an underlying theory and takes the integer value.
In fact, we find that the partition function is not modified from the gauged WZW model  by the numerical simulation:
\begin{eqnarray}
Z_{\text{GWZWM}}^{SU(N)}(T^2,k,t) = \frac{(N+k-1)!}{(N-1)! k!}.
\label{genus 1 GWZWM}
\end{eqnarray}

Next, we investigate the partition function of genus-$0$, sphere.
By the numerical simulation, we conjecture that the partition function behaves as
\begin{eqnarray}
Z_{\rm GWZWM}^{SU(N)} (S^2,k,t)
= \frac{1}{\prod_{a=1}^N (1-t^a)}.
\label{genus 0 GWZWM}
\end{eqnarray}
Notice that this does not depend on the level $k$
and of course coincides with the partition function of the gauged WZW model in the limit $t\rightarrow 0$.

\begin{table}[t]
\begin{center}
\begingroup
\renewcommand{\arraystretch}{1.3}
  \begin{tabular}{|c|c|c|c|} \hline
  Genus & $L=k$ & $M=N$ & Partition Function \\ \hline\hline
 2  & 2 & 3 & $(1-t)^3   (45 + 99 t + 99 t^2 + 45 t^3)$ \\ 
   & 3 & 3 & $(1-t)^3    (166 + 332 t + 252 t^2 + 86 t^3 + t^4)$ \\ 
   & 4 & 3 & $(1-t)^3 (504 + 810 t + 396 t^2 + 126 t^3 + 36 t^4)$\\
   & 5 & 3 & $(1-t)^3 (1332 + 1512 t + 369 t^2 + 243 t^3 + 144 t^4)$\\ 
   & 2 & 4 & $4 (1-t)^4 (1+t)^2 (35 + 50 t + 86 t^2 + 50 t^3 + 35 t^4 )$ \\
  3 & 3 & 2 & $8 (1 - t)^4 (3 + 2 t) (5 + 4 t)$ \\
   & 4 & 2 & $(1-t)^4(329+280 t +86 t^2+ + 8 t^3 + t^4)$\\ 
   & 2 & 3 & $27 (1-t)^6 (1 + t)^2 (3 + 4 t + 3 t^2) (5 + 6 t + 5 t^2)$\\ 
  4 & 3 & 2 & $16 (1 - t)^6 (2 + t) (5 + 4 t)^2$\\
  5 & 3 & 2 & $ 32 (1 - t)^8 (5 + 4 t)^2 (7 + 8 t + 2 t^2)$ \\ \hline
    \end{tabular}
    \endgroup
    \end{center}
\caption{The partition function  of the $SU(N)/SU(N)$ gauged WZW-matter model with the level $k$ on the   genus-$h$ Riemann surface.} 
\label{table:GWZWM other}
\end{table}

In the case of  genus-$h$ ($h \ge2$), we can not conjecture how the partition function behaves in arbitrary $k$ and $N$.
Thus, we consider two special cases: $N=2$, $k=\text{arbitrary}$, $h=2$ and $N=k=2$, $h=\text{arbitrary}$.
We list the result in the former  and  later case at Table \ref{table:su(2) GWZWM genus 2} and
Table \ref{table:su(2) GWZWM higher genus}, respectively.

In the former case, we conjecture that from Table \ref{table:su(2) GWZWM genus 2} the partition function  behaves as
\begin{eqnarray}
Z_{\rm GWZWM}^{SU(2)}{(\Sigma_2, k,t)} &=&
 (1-t)^2\left( \frac{(k+3)(k+2)(k+1)}{6}\right.\nonumber\\
 && \left.\quad -\frac{(k-7)k(k+1)}{3}t + \frac{(k-3)(k-2)(k-1)}{6}t^2\right).
\end{eqnarray}

In the later case, we also conjecture that from Table \ref{table:su(2) GWZWM higher genus} the partition function  behaves as
\begin{eqnarray}
Z_{\rm GWZWM}^{SU(2)}{(\Sigma_h, k=2,t)} = 2^{h-1}(2^h + 1) (1-t)^{2h-2}(1+t)^{h-1}.
\end{eqnarray}
We can not conjecture the  general form of the partition functions in the other cases
but list the result of  several cases at Table \ref{table:GWZWM other}.
As see Table  \ref{table:su(2) GWZWM genus 2}, Table \ref{table:su(2) GWZWM higher genus} and Table \ref{table:GWZWM other}, 
we find that all the coefficients for the power of $t$ in the partition function are integer.
The partition function itself changes but this property does not change, even if we change the normalization  
such that the partition function of the gauged WZW-matter model becomes that of the gauged WZW model in the limit $t\rightarrow 0$.
This implies that the partition function is a topological invariant.
Furthermore, the partition function of the $U(N)/U(N)$ gauged WZW-matter model also has same property.

\subsection{Gauge/Bethe correspondence}
\label{subsec:GWZWM Gauge/Bethe}

In this subsection, we are going to establish the Gauge/Bethe correspondence between the $U(N)/$ $U(N)$ or $SU(N)/SU(N)$ gauged WZW-matter model and the $q$-boson model.

First of all, let us see that the localized configurations in the $U(N)/U(N)$ gauged WZW-matter model coincide with the Bethe Ansatz equations in the $q$-boson model.
We change the parameterization of the coupling constant $t$ as $t = e^{-2\pi \zeta}$ in the localized configurations (\ref{localization configuration})
in order to see manifest coincidence with the Bethe Ansatz equations (\ref{log Bethe equation q-boson}) in the $q$-boson model.
Then, we can rewrite (\ref{localization configuration}) as
\begin{eqnarray}
2\pi i k x_j =  2\pi i I_j + \sum_{k=1}^N \log\frac{\sin[\pi( i\zeta + (x_j-x_k))]}{\sin[\pi( i\zeta -(x_j-x_k))]}
\label{localization configuration2}
\end{eqnarray}
where $I_j$ is (half-)integers when $N$ is (even) odd.
We identify the level $k$, the dual Coxeter number $N$ of $\mathfrak{u}(N)$ and the coupling constant $\zeta$ in the $U(N)/U(N)$ gauged WZW-matter model
with the total site number $L$, the particle number $M$ and the coupling constant $\eta$ in the $q$-boson model, respectively
\footnote{Note that these identifications of the parameters are different from the ones in \cite{Okuda:2012nx}. 
In \cite{Okuda:2012nx}, we  investigated the relations between the $U(N)/U(N)$ or $SU(N)/SU(N)$ gauged WZW model, and the phase model under $k\equiv M$ and $N\equiv L$.
This is because  the WZW primary fields and the modular matrix in the $SU(N)$ WZW model completely coincide with the Bethe roots  and the norm between the eigenstates in the phase model, respectively.
Therefore, the identification $k \equiv M$ and $N \equiv L$ in the case of \cite{Okuda:2012nx} is more natural than  $k \equiv L$ and $N \equiv M$.
However, all models do not have  invariance under the level-rank duality transformation.
In fact, such transformation is unlikely to exist in the  $G/G$ gauged WZW-matter model.
}.
Moreover, we identify the Cartan part $\{\phi_1,\cdots,\phi_N\}$ of the field $g$ in the gauged WZW-matter model as the Bethe roots $\{x_1,\cdots,x_N\}$ in the $q$-boson model.
Then, we find that the localized configurations (\ref{localization configuration2}) in the gauged WZW-matter model coincide with the Bethe Ansatz equations (\ref{log Bethe equation q-boson}) in the $q$-boson model
under these identifications.

Next, let us investigate the relation between the set of  the piecewise independent solutions of the Bethe Ansatz equations for the $q$-boson model and the set $\{{\rm Sol}\}$ of $\{x_1,\cdots,x_N\}$ which contributes to the partition function of the gauged WZW-matter model.
It is necessary for the Bethe states to form a complete system  that the number of the piecewise independent solutions of the Bethe Ansatz equations for the $q$-boson model is $(N+k -1)!/(N-1)!k!$.
Although it is  nontrivial whether this number coincides with the number of elements of the set $\{{\rm Sol}\}$,
we can numerically confirm it.
That is to say, the number of the elements of the set $\{{\rm Sol}\}$  is $(N+k -1)!/(N-1)!k!$
and coincides with the number of the piecewise independent solutions of the Bethe Ansatz equations for the $q$-boson model.
This circumstance is equal to  that of the correspondence between the $U(N)/U(N)$ gauged WZW model and the phase model.
Thus, we  have established  equivalence between $\{\rm Sol\}$ and the set of the independent solutions of the Bethe Ansatz equations for the $q$-boson model.

Finally, we consider the partition function for the $U(N)/U(N)$ gauged WZW-matter model.
Under the above identifications, the norm between the eigenstates of the Hamiltonian in the $q$-boson model (\ref{q-boson Bethe norm}) becomes
\begin{eqnarray}
\langle \psi(\{e^{2\pi ix}\}_N)|\psi(\{e^{2\pi ix}\}_N)\rangle
= \frac{\prod_{a,b=1}^N (e^{2\pi ix_a} t - e^{2\pi ix_b})}{\prod_{\substack{a,b=1\\a\neq b}}^N (e^{2\pi ix_a} -e^{2\pi ix_b})}
\cdot\det_N\Phi_{a,b}^{\prime}(\{x\}_N).
\label{q-boson Bethe norm2}
\end{eqnarray}
Here, the Gaudin matrix (\ref{q-boson Gaudin})
becomes 
\begin{eqnarray}
\Phi^{\prime}_{a,b} (\{e^{2\pi i x}\}_N) 
&=&\delta_{a,b} \left\{-k e^{-2\pi i x_b}
+ \sum_{c=1}^N\frac{(t^2 - 1)e^{2\pi ix_c}}{(t e^{2\pi ix_a} -e^{2\pi ix_c})(t e^{2\pi ix_c}  -e^{2\pi ix_a})}\right\}\nonumber\\
&&\hspace{2cm}- \frac{(t^2 - 1)e^{2\pi ix_a}}{(t e^{2\pi ix_a} -e^{2\pi ix_b})(t e^{2\pi ix_b}  -e^{2\pi ix_a})}.
\label{q-boson Gaudin2}
\end{eqnarray}
Thus, it is obvious that the partition function (\ref{normalized partition U(N)})  
is expressed by the summation  of the norms in terms of all the eigenstates:
\begin{eqnarray}
Z_{\rm GWZWM}^{U(N)}(\Sigma_h,k,t) 
= \left(\frac{N+k}{k}\right)^h \sum_{x_1,\cdots,x_N\in\{{\rm Sol}\}}\langle \psi(\{ e^{2\pi ix}\}_N)|\psi(\{e^{2\pi ix}\}_N)\rangle^{h-1}.
\label{partition q-boson}
\end{eqnarray}
As a result, we find that the $U(N)/U(N)$ gauged WZW-matter model corresponds to the $q$-boson model in a sense of the Gauge/Bethe correspondence.

Similarly, we can obtain the following expression for  the partition function (\ref{normalized partition SU(N)}) of the $SU(N)/SU(N)$ gauged WZW-matter model:
\begin{eqnarray}
Z_{\rm GWZWM}^{SU(N)}(\Sigma_h,k,t) 
= \left(\frac{N}{k}\right)^h \sum_{x_1,\cdots,x_N\in\{{\rm Sol}\}}\langle \psi(\{ e^{2\pi ix}\}_N)|\psi(\{e^{2\pi ix}\}_N)\rangle^{h-1}.
\label{partition q-boson SU(N)}
\end{eqnarray}
This circumstance is also equal to  the correspondence between the gauged WZW model and the phase model.
Thus, we find that the $SU(N)/SU(N)$ gauged WZW-matter model also corresponds to the $q$-boson model  in a sense of the Gauge/Bethe correspondence.
This correspondence is a one parameter deformation of the  correspondence between the $SU(N)/SU(N)$ or $U(N)/U(N)$ gauged WZW model and the phase model. 
In next subsection, we will consider a reason why the  Gauge/Bethe correspondence between the $SU(N)/SU(N)$ gauged WZW-matter model and the $q$-boson model works well from the viewpoint of the axiom of the topological quantum field theory.

\subsection{Partition function from the commutative Frobenius algebra}
\label{subsec:GWZWM Axiom}

In this subsection, we study the partition function of the $SU(N)/SU(N)$ gauged WZW-matter model
from the viewpoint of the axiomatic system of the TQFT.
It is well known that the TQFT has the axiomatic formulation  given by Atiyah \cite{Atiyah:1989vu} and Segal \cite{Segal:2002ei}.
In particular, the category of two dimensional TQFTs is equivalent to the category of commutative Frobenius algebras.
See \cite{Dijkgraaf:1997ip, Dijkgraaf, Kock:book} for details.

Recently, Korff constructed a new commutative Frobenius algebra from the $q$-boson model \cite{Korff:2013rsa}.
Since the $q$-boson model also appears in the $SU(N)/SU(N)$ gauged WZW-matter model as discussed previous section,
it  is natural  that the $SU(N)/SU(N)$ gauged WZW-matter model is related to the commutative Frobenius algebra constructed from the $q$-boson model.
We can actually show that the partition function of  the $SU(N)/SU(N)$ gauged WZW-matter model coincides with the partition function of the commutative Frobenius algebra constructed from the $q$-boson model up to the overall factor.
This implies that  the $SU(N)/SU(N)$ gauged WZW-matter model can be regarded as a Lagrangian realization of the commutative Frobenius algebra  constructed by Korff.

From here, we briefly summarize necessary ingredients in \cite{Korff:2013rsa} to show the agreement between the partition function of the both theories.
We first explain a theorem (Theorem 7.2 in \cite{Korff:2013rsa}) without the proof that a commutative Frobenius algebra can be constructed on the $N$-particle subspace of the Fock space in the $q$-boson model
\footnote{Note that we now interchange $k$ with $N$ for results in \cite{Korff:2013rsa}.}.

We give several definitions of ingredients in the theorem as preparation.
Let $\mathcal{P}^{+}_{N}$ be a set of dominant integrable (positive) weights of $\mathfrak{gl}(N)$ and  $\mathcal{A}^{+}_{N,k}$ be a subset of  $\mathcal{P}^{+}_{N}$
defined by
\begin{eqnarray}
\mathcal{A}^{+}_{N,k} =\{ (\mu_1, \mu_2, \cdots, \mu_N) \in \mathcal{P}^{+}_{N} ~|~ k \ge \mu_1 \ge \mu_2 \ge \cdots \ge \mu_k \ge 1 \}.
\end{eqnarray}
Then, $\mathcal{A}^{+}_{N,k}$ one-to-one corresponds to  the set of the independent solutions of Bethe Ansatz equations  for the $q$-boson  model:
\begin{eqnarray}
\lambda_j^k = \prod_{\substack{k=1\\k\neq j}}^N \frac{\lambda_j t -\lambda_k}{\lambda_j -\lambda_k t}\quad\text{for}\quad j=1,\cdots,N
\end{eqnarray}
where $0\le t <1$.
This set is also in one-to-one correspondence with $\{\rm Sol\}$  defined in the previous subsection.

Next, we define a Bethe vector and its dual vector as
\begin{eqnarray}
|\Psi_{\sigma}\rangle = \prod_{j=1}^N B\left((\lambda_{\sigma})_j^{-1}\right)|0 \rangle\quad\text{and}\quad
\langle \Psi^*_{\sigma}| = \frac{1}{||\Psi_{\sigma}||^2}\langle 0| \prod_{j=1}^NC\left((\lambda_{\sigma})^{-1}_j\right),
\end{eqnarray}
respectively, where $\lambda_{\sigma}$ denotes a Bethe root corresponding to a partition $\sigma$.
Here,  $||\Psi_{\sigma}||^2$ 
is defined by
\begin{eqnarray}
||\Psi_{\sigma}||^2 = \langle 0|\prod_{j=1}^N C\left((\lambda_{\sigma})^{-1}_j\right) \prod_{j=1}^N B\left((\lambda_{\sigma})^{-1}_j\right)|0\rangle.
\end{eqnarray}
Therefore, we have the identity $\langle \Psi_{\mu}^*|\Psi_{\nu}\rangle = \delta_{\mu\nu}$.

Furthermore, we define a vector $|\mu\rangle, \mu \in {\cal A}_{N,k}^+$ in the $N$-particle subspace of the Fock space ${\cal F}_N^{\otimes k}$ in the $q$-boson model as
\begin{eqnarray}
|\mu\rangle = |\mu_1\rangle \otimes |\mu_2\rangle\otimes \cdots \otimes |\mu_N\rangle.
\end{eqnarray}
Then,  we define 
 a transition matrix $S_{\mu\nu}(t)$ from the basis of normalized Bethe vectors $\{ |\Psi_{\nu}\rangle : \nu\in{\cal A}_{N,k}^+\}$  
to the vector $\{|\mu\rangle : \mu\in{\cal A}_{N,k}^+\}$ in the Fock space ${\cal F}_{N}^{\otimes k}$   by
\begin{eqnarray}
S_{\mu\nu}(t) = ||\Psi_{\nu}|| \langle \Psi^{\ast}_{\nu} | \mu \rangle. 
\label{deformedS}
\end{eqnarray}
It is also shown in \cite{Korff:2013rsa} that the transition matrix satisfies the following relation 
\begin{eqnarray}
S_{\mu \lambda}^{-1}(t)=b_{\lambda}(t) S_{\lambda^* \mu}(t).
\label{inverseS}
\end{eqnarray}
where  $\ast$-involution on $\mu$ in ${\cal A}^+_{N,k}$ is defined as
\begin{eqnarray}
\mu^*_i= 
\left\{
\begin{array}{c}
k-\mu_{N-i+1} \quad (k-\mu_{N-i+1} \neq 0) \\
\hspace{0.75cm}
 k \hspace{1.4cm} (k-\mu_{N-i+1} =0)
\end{array}
\right.
\end{eqnarray}
for $i =1,\cdots,N$ and 
  $b_{\mu}(t)$ is defined by
\begin{eqnarray}
b_{\mu}(t)=\prod_{i \ge 1} \prod_{j=1}^{m_i(\mu)} (1-t^j). 
\end{eqnarray}
Here, $m_i(\mu)$ is the  multiplicity of $i$ in $\mu$ and is defined by $m_i(\mu) = \mathrm{Card}\{ j : \mu_j =i \}$.

Let us label the partition $k^N$ with ``0".
When we set $\mu = 0$ in (\ref{deformedS}), the transition function $S_{0 \nu}(t)$ is expressed by
\begin{eqnarray}
S_{0 \nu}(t)=\frac{1}{||\Psi_{\nu}||}.
\end{eqnarray}

Korff proved that a commutative Frobenius algebra can be constructed on the $N$-particle subspace of the Fock space ${\cal F}_N^{\otimes k}$ in the $q$-boson model, and asserted a following theorem:
\begin{theorem}[Commutative Frobenius algebra \cite{Korff:2013rsa}]
Let $\mathbbm{k}$ be the algebraically closed field of the Puiseux series 
and  $\mathfrak{F}_{k,N} := {\cal F}_N^{\otimes k} \otimes_{\mathbb{C} (t)} \mathbbm{k}$.
Define for $ \mu, \nu \in {\cal A}_{N,k}^+$ the product
\begin{eqnarray}
|\mu \rangle \circledast |\nu \rangle :=\sum_{\rho \in \mathcal{A}^{+}_{N,k}} N^{\rho}_{\mu \nu}(t) |\rho \rangle 
\end{eqnarray}
where the structure constant of the commutative Frobenius algebra $N_{\mu\nu}^{\lambda}(t)$ is defined as
\begin{eqnarray}
N^{\lambda}_{\mu \nu}(t)=\sum_{\sigma \in \mathcal{A}^{+}_{N,k} } \frac{ S_{\mu \sigma}(t) S_{\nu \sigma}(t)  S^{-1}_{\sigma \lambda}(t) }{S_{0 \sigma}(t)}.
\label{defVerlinde}
\end{eqnarray}
Here, the transition matrix $S_{\mu\nu}(t)$ is defined in (\ref{deformedS}).

Moreover, define the associative, nondegenerate bilinear form  $\eta : \mathfrak{F}_{k,N} \otimes \mathfrak{F}_{k,N} \to \mathbbm{k}$
\begin{eqnarray}
\eta(|\mu \rangle \otimes |\nu \rangle )=\eta_{\mu \nu}(t):=\frac{\delta_{\mu \nu^{*}}}{b_{\nu}(t)}.
\label{bilinear}
\end{eqnarray}
Then, $(\mathfrak{F}_{k,N}, \circledast, \eta)$ is a commutative Frobenius algebra with a unit $|k^N\rangle, k^N=(k, k, \cdots, k)$.
\end{theorem}

From now on, we investigate relations between the partition function of the $SU(N)/SU(N)$ gauged WZW-matter model and 
the partition function of the TQFT equivalent to the commutative Frobenius algebra.
Recall that the partition function of the  $SU(N)$/$SU(N)$ gauged WZW-matter model is expressed
by the summation of the norms between the eigenvectors in the $q$-boson model in terms of all the Bethe roots, (\ref{partition q-boson SU(N)}).
Then, we find that the partition function  (\ref{normalized partition SU(N)}) can be rewritten by using the transition matrix $S_{0\mu}(t)$ as  
 \begin{eqnarray}
Z_{\rm GWZWM}^{SU(N)}(\Sigma_h, k, t)=\left(\frac{N}{k}\right)^h \sum_{\mu \in \mathcal{A}^{+}_{N,k} }  \frac{1}{S^{2h-2}_{ 0 \mu}(t)}.
\label{partitionGWZWMmod}
\end{eqnarray}
We can show this formula by  the fact that  the set of the independent Bethe roots $\{\rm Sol\}$ one-to-one corresponds to ${\cal A}_{N,k}^+$
and by a following identity:
\begin{eqnarray}
\sum_{\sigma \in {\cal A}_{N,k}^+} \left\{\langle 0|\prod_{j=1}^N C((\lambda_{\sigma})_j) \prod_{j=1}^N B((\lambda_{\sigma})_j)|0\rangle\right\}^Z
= \sum_{\sigma \in {\cal A}_{N,k}^+} ||\Psi_{\sigma}||^{2Z}
\end{eqnarray}
for $Z\in \Z$.
Here, we have used the explicit expression for the norm in the $q$-boson model (\ref{q-boson Bethe norm}) to prove this identity.

\begin{figure}[t]
\begin{center}
\includegraphics[width = 4cm]{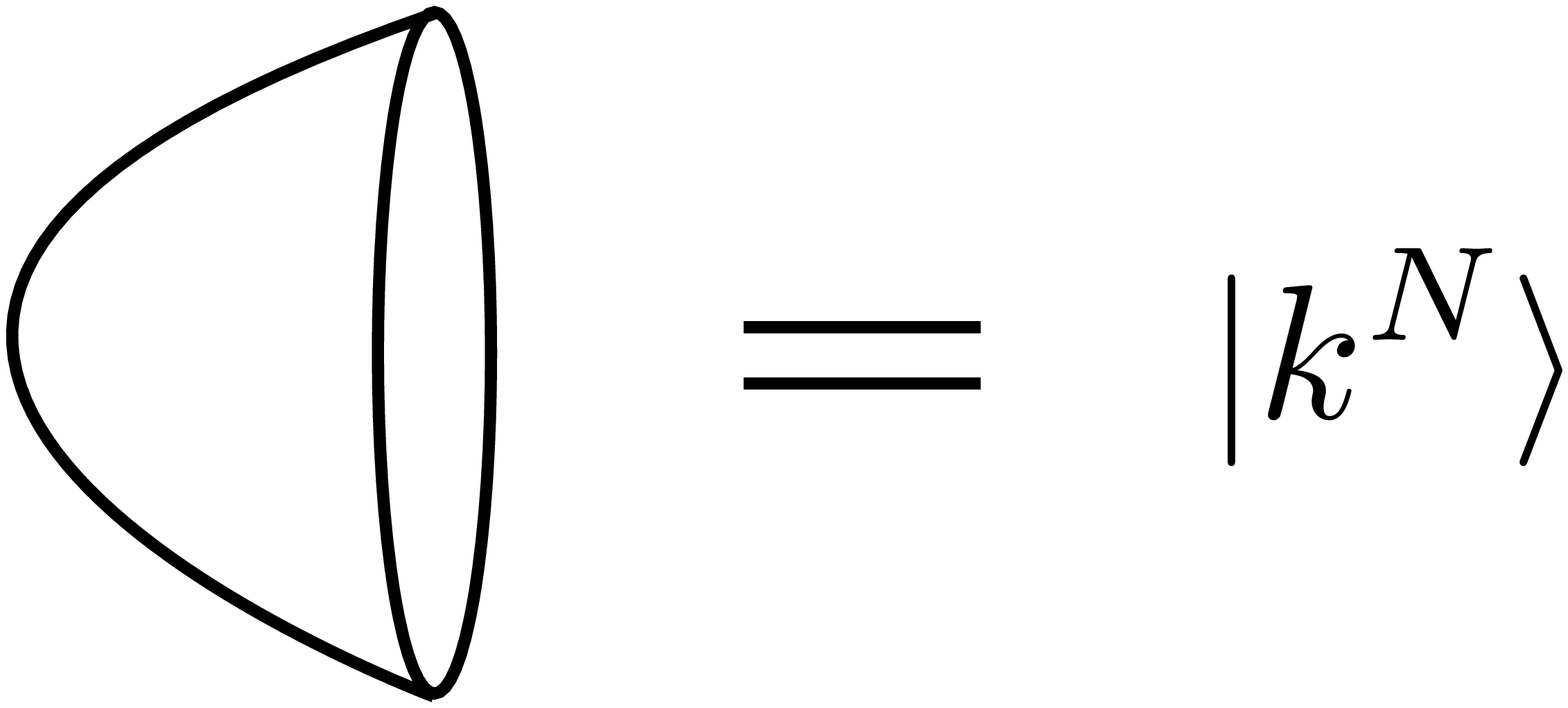}
\caption{ Unit}
\label{fig:disc1}
\end{center}
\vspace{2mm}
 \begin{minipage}{10cm}
  \begin{center}
   \includegraphics[width=4.5cm]{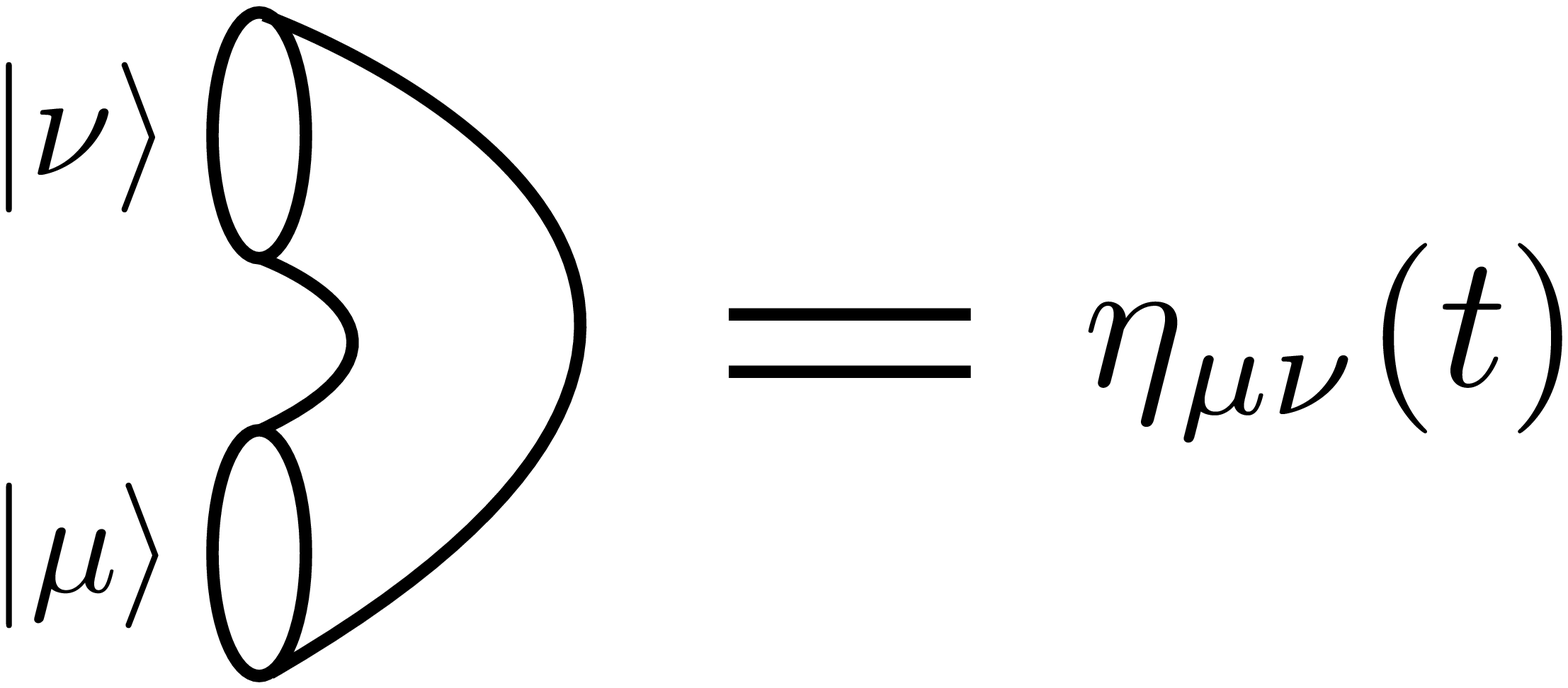}
  \end{center}
 \end{minipage}
 \begin{minipage}{0cm}
  \begin{center}
   \includegraphics[width=4.5cm]{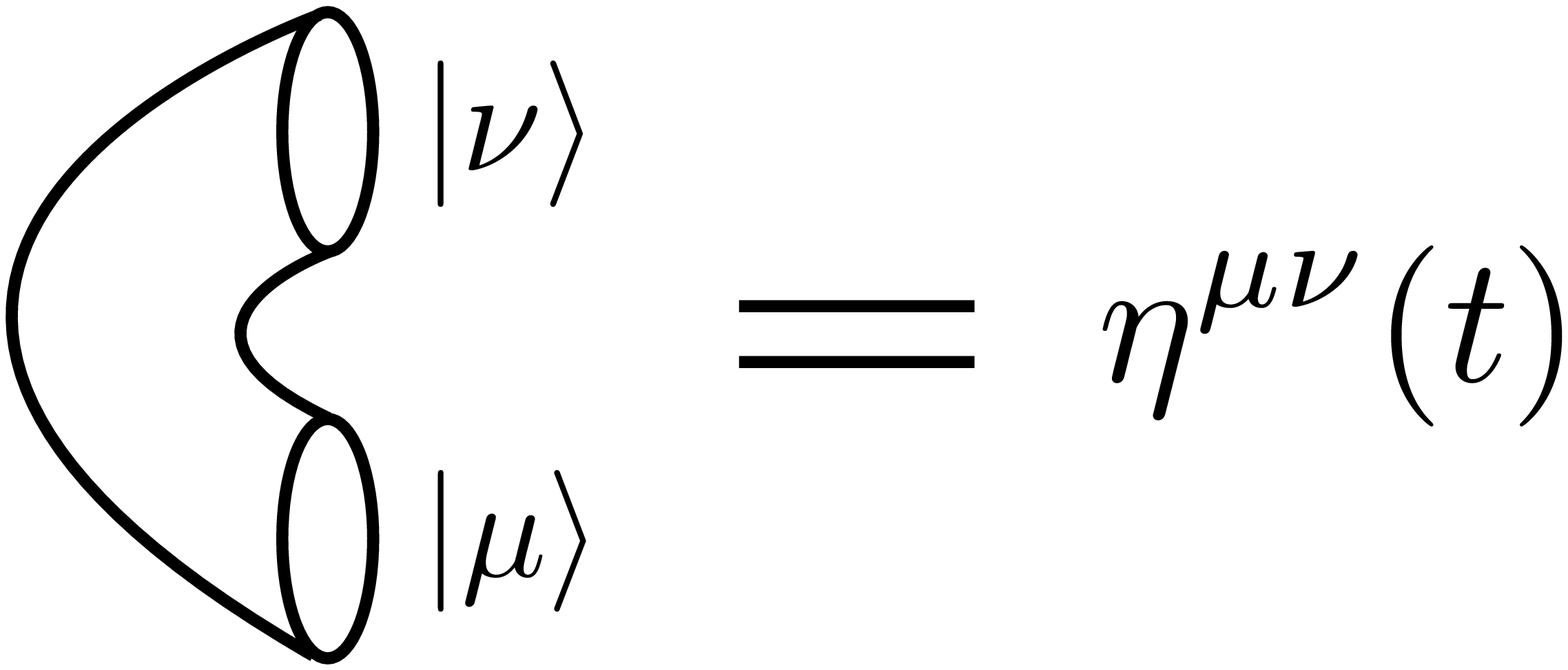}
  \end{center}
 \end{minipage}
 \caption{Nondegenerate bilinear forms}
 \label{fig:bilinear form}
\vspace{2mm}
\begin{center}
\includegraphics[width= 7cm]{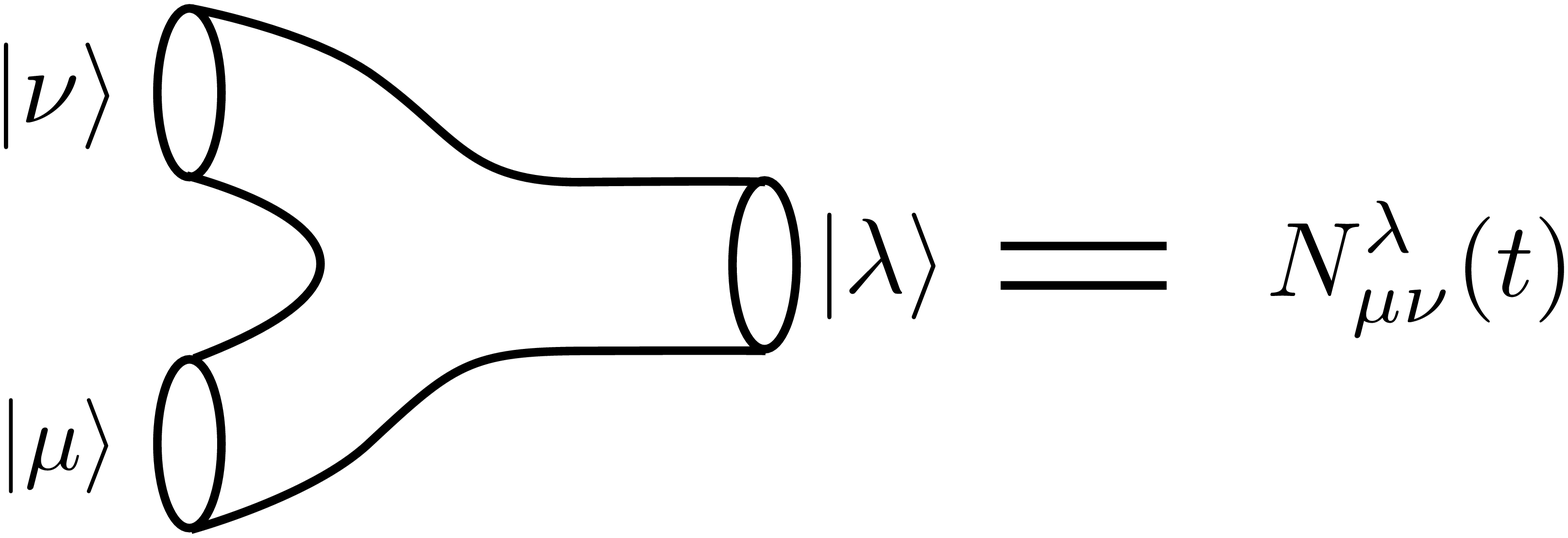}
\caption{Structure constant}
\label{fig:pants}
\end{center} 
\end{figure}
On the other hand, we construct the partition function from the commutative Frobenius algebra.
 In order to do this, let us graphically represent the building blocks of the commutative Frobenius algebra, that is to say, the unit $|k^N\rangle$, the nondegenerate bilinear form $\eta_{\mu\nu}(t)$ and the structure constant $N_{\mu\nu}^{\lambda}(t)$.
We first assign the unit  $| k^N \rangle $ to a disc with an outgoing boundary (Figure \ref{fig:disc1}).  
Secondly, we assign the non-degenerate bilinear form $\eta_{\mu \nu}(t)$ and its inverse $\eta^{\mu\nu}(t)$ to a cylinder with  two incoming boundaries (the left picture of Figure \ref{fig:bilinear form}) and to a cylinder with two outgoing boundaries (the right picture of Figure \ref{fig:bilinear form}), respectively.
Finally, we assign the structure constant  (\ref{defVerlinde}) to a sphere with three boundaries (Figure \ref{fig:pants}).

\begin{figure}[t]
\begin{center}
\includegraphics[width=6cm]{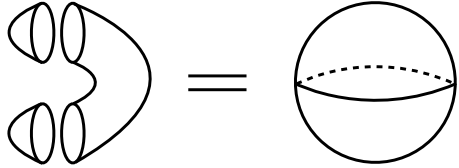}
\caption{Partition function on a sphere $S^2$}
\label{fig:sphere}
\end{center}
\end{figure}

Let us  construct the partition function of the commutative Frobenius algebra on the genus-$h$ Riemann surface by gluing  the surfaces.
First of all, we consider the case of the genus-$0$.
In the genus-$0$, we can  construct the partition function by gluing  two outgoing discs and a cylinder with two incoming boundaries  like Figure \ref{fig:sphere}.
Therefore, we find that the partition function agrees with the partition function of the $SU(N)/SU(N)$ gauged WZW-matter model on $S^2$:
\begin{eqnarray}
Z^{N,k}_{\mathrm{TQFT}}(S^2)=\eta_{k^N, k^N}(t) =\frac{\delta_{k^N, k^N}}{b_{k^N}(t)}=\prod_{i=1}^N\frac{1}{(1-t^i)}.
\end{eqnarray}

\begin{figure}[t]
\begin{center}
\includegraphics[width=6cm]{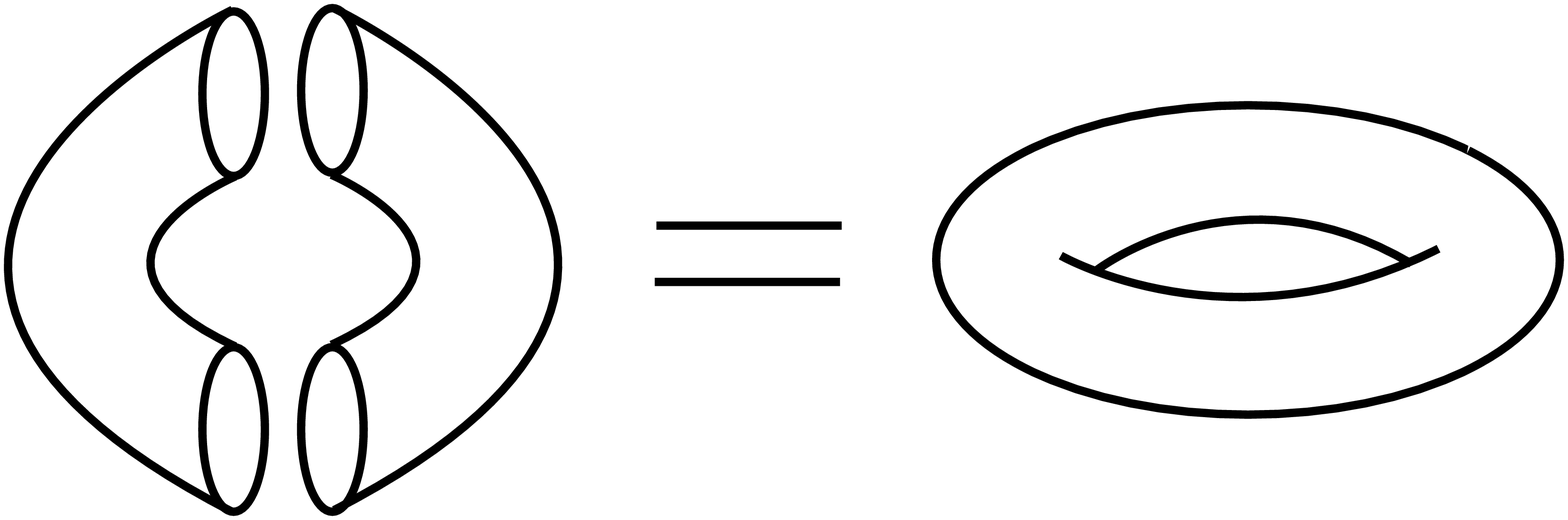}
\caption{Partition function on a torus $T^2$}
\label{fig:torus}
\end{center}
\end{figure}

\begin{figure}[t]
\begin{center}
\includegraphics[width=8cm]{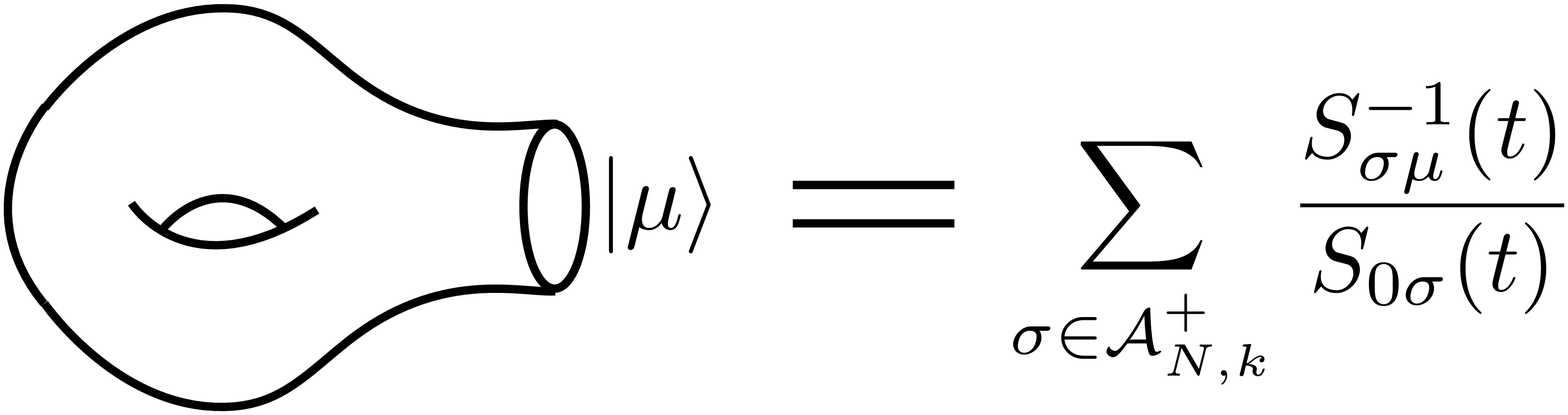}
\caption{Handle operator}
\label{fig:oneholded}
\end{center}
\end{figure}

\begin{figure}[t]
\begin{center}
\includegraphics[width=10cm]{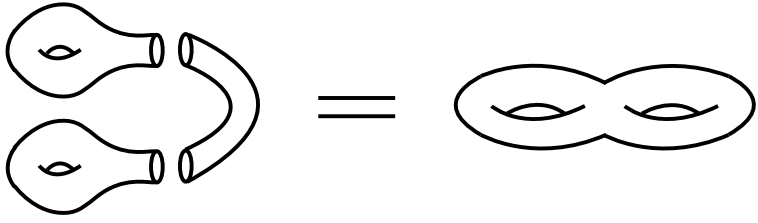}
\caption{Partition function on a genus-$2$ Riemann surface}
\label{fig:2torus}
\end{center}
\end{figure}

Next, we consider the case of the genus-$1$.
In this case, the partition function can be constructed by gluing a cylinder with two incoming boundaries and two outgoing boundaries like Figure \ref{fig:torus},
and be therefore expressed by
\begin{eqnarray}
Z^{N,k}_{\mathrm{TQFT}}(T^2)
=\sum_{\mu, \nu \in \mathcal{A}^{+}_{N,k}} \eta_{\mu  \nu}(t) \eta^{\mu  \nu} (t)
=\mathrm{dim}\hspace{0.5mm}\mathcal{A}^{+}_{N, k}
=\frac{(k+N-1)!}{(k-1)! N!}.
\end{eqnarray}
Thus, we find that  the partition function of the $SU(N)/SU(N)$ gauged WZW-matter model on the torus  coincides with  that of the commutative Frobenius algebra  up to the overall factor:
\begin{eqnarray}
Z^{SU(N)}_{\text{GWZWM}}(T^2,k,t)
=\frac{N}{k}Z^{N,k}_{\mathrm{TQFT}}(T^2).
\end{eqnarray}
Similarly, we can construct the partition function of the commutative Frobenius algebra on the higher genus Riemann surface.
In order to construct this, we introduce a handle operator, the torus with one puncture in Figure \ref{fig:oneholded}.
This can be constructed by gluing  a cylinder with two outgoing boundaries, and a sphere with one outgoing and two incoming boundary boundaries.
Thus, we obtain 
\begin{eqnarray}
\sum_{ \nu, \rho \in \mathcal{A}^+_{N, k}} \eta^{ \nu \rho}(t) N^{\mu}_{\nu \rho}(t)
&=&
\sum_{ \nu, \rho, \sigma \in \mathcal{A}^+_{N, k}} b_{\nu}(t) \delta^{ \nu^* \rho} 
\frac{S_{\nu \sigma}(t) S_{\rho \sigma}(t) S^{-1}_{ \sigma \mu}(t)}{S_{0 \sigma}(t)}
\nonumber \\
&=&
\sum_{ \nu, \sigma \in \mathcal{A}^+_{N, k}}   \frac{S^{-1}_{ \sigma \nu}(t) S_{\nu \sigma}(t) S^{-1}_{ \sigma \mu}(t)}{S_{0 \sigma}(t)}
\nonumber \\
&=&
\sum_{ \sigma \in \mathcal{A}^+_{N, k}}   \frac{ S^{-1}_{ \sigma \mu}(t)}{S_{0 \sigma}(t)}.
\end{eqnarray}
Here, we  have used (\ref{inverseS}) from the first line to the second line.
By using this handle operator, we can construct the partition function on the higher genus Riemann surface.
For example, the  partition function on the genus-$2$ Riemann surface constructed like Figure \ref{fig:2torus} becomes
\begin{eqnarray}
Z^{N,k}_{\rm TQFT}(\Sigma_2) &=& \sum_{\mu, \nu,  \rho, \sigma \in \mathcal{A}^+_{N, k}}  
 \frac{ S^{-1}_{ \sigma \mu}(t)}{S_{0 \sigma}(t)} \frac{S^{-1}_{ \rho \nu}(t)}{ S_{0 \rho}(t)} \frac{\delta_{\mu \nu^*}}{b_{\nu}(t)} \nonumber \\
&=&\sum_{\nu,  \rho, \sigma \in \mathcal{A}^+_{N, k}}  
 \frac{ S_{ \nu  \sigma }(t)}{S_{0 \sigma}(t)} \frac{S^{-1}_{ \rho \nu}(t)}{ S_{0 \rho}(t)}  \nonumber \\
&=& 
\sum_{\mu \in \mathcal{A}^{+}_{N,k}}  \frac{1}{S^{2}_{ 0 \mu}(t)}.
\end{eqnarray}
The  partition function on the genus-$h$ Riemann surface can  be recursively constructed in the similar manner with the case of the genus-$2$, and
is expressed by
\begin{eqnarray}
Z^{N,k}_{\mathrm{TQFT}}(\Sigma_h) 
= \sum_{\mu \in \mathcal{A}^{+}_{N,k}}  \frac{1}{S^{2h-2}_{ 0 \mu}(t)}.
\end{eqnarray} 
As compared with (\ref{partitionGWZWMmod}), we find that the partition function of the commutative Frobenius algebra coincides with that of the $SU(N)/SU(N)$ gauged WZW-matter model up to the overall factor:
\begin{eqnarray}
Z_{\rm GWZWM}^{SU(N)}(\Sigma_h, k, t)
=\left(\frac{N}{k}\right)^h Z^{N,k}_{\rm TQFT}(\Sigma_h).
\label{genus-h GWZWM and cFA}
\end{eqnarray}
Thus, we have clarified the equivalence between the partition function of the $SU(N)/SU(N)$ gauged WZW-matter model 
and  the two dimensional TQFT equivalent to the commutative Frobenius algebra constructed by Korff.
In fact, we can concretely check the relation (\ref{genus-h GWZWM and cFA}) by using an algorithm given by  section 7.3.1 in \cite{Korff:2013rsa}, which calculates the structure constant of the commutative Frobenius algebra from the structure constants of the (restricted) Hall algebra.
In several cases, we have verified  the agreement with the numerical results obtained in the previous section.
Therefore,  the $SU(N)/SU(N)$ gauged WZW-matter model can be regarded as a Lagrangian realization of the commutative Frobenius algebra  constructed by Korff.
We conjecture that the Gauge/Bethe correspondence works well only if a commutative Frobenius algebra can be constructed from a certain integrable system,
just as  \cite{Korff:2013rsa, Korff:2010},
and that this is the underlying mathematical structure of the Gauge/Bethe correspondence.

The Gauge/Bethe correspondence means that a certain topological gauge
theory have a hidden quantum integrable structure.
In the Gauged WZW-matter model/$q$-boson model correspondence, we have identified the dual Coxeter number and the
level in the gauged WZW-matter model as the total particle number and the
total site number in the $q$-boson model, respectively.
This implies that the whole collection of the $SU(N)/SU(N)$ gauged WZW-matter models with the different
ranks has the quantum integrable structure of the  $q$-boson model.
Since this integrable structure relates  different topological quantum field theories with different
ranks, it looks strange but may be interesting property.
In particular, it would be interesting to investigate
the roles of the Yang-Baxter algebra in topological field theories. The
Yang-Baxter algebra in the $q$-boson model controls the gauged WZW-matter
model. That is to say, the operators $B(\lambda)$ and $C(\lambda)$ in the Yang-Baxter
algebra whose spectral parameters satisfy the Bethe Ansatz equations,
create and annihilate the fields in the collections of the gauged WZW-matter models, because we
identify the Fock space of the $q$-boson model as the space of the integrable
weight with the fields in the $SU(N)/SU(N)$ gauged WZW-matter model.

\subsection{Correlation functions of the gauged WZW-matter model}
\label{sec:Correlation}
We can easily calculate the correlation functions of BRST-closed operators
from the viewpoint of the cohomological localization.
In this subsection, we investigate a question
how the correlation functions of the field $g$ in the $SU(N)/SU(N)$ gauged WZW-matter model are related to quantities in the $q$-boson model.

For simplicity, we first consider a one-point function. 
The generalization to correlation functions is straightforward.
The one-point function of $g^r$ with a positive integer number $r$ is defined as
\begin{eqnarray}
&&\hspace{-1cm} \langle \Tr g^r \rangle(\Sigma_h,SU(N),k, t)\nonumber\\
&& \hspace{-1cm}\quad = \int \D g  \D^2 A \D^2 \lambda \D\Phi \D\Phi^{\dagger} \D\psi \D\psi^{\dagger} \D^2 \varphi \D^2 \chi ~\Tr g^{r}
e^{-k S_{\rm GWZWM}(g, A, \lambda, \Phi, \Phi^{\dagger}, \psi, \psi^{\dagger}, \varphi, \chi) }.
\end{eqnarray}
We apply the cohomological localization to this one-point function as with the case of the partition function in section \ref{subsec:GWZWM localization}.
From the viewpoint of the cohomological localization, we find that the localized configurations (\ref{localization configuration}) do not change and the one-point function becomes
\begin{eqnarray}
&&\hspace{-1cm}\langle \Tr g^r \rangle(\Sigma_h,SU(N),k,t)
\nonumber\\
&&\hspace{-1cm} \quad = \left(\frac{N}{k}\right)^h  \sum_{\{\phi_1,\cdots,\phi_N\}\in\{\rm Sol\}}
\sum_{c=1}^Ne^{2\pi i r  \phi_c}
\left\{(1-t)^N \mu_q(\phi) 
 \prod_{\substack{a,b=1\\a \neq b}}^N
 \frac{e^{2\pi i\phi_a} - t e^{2\pi i\phi_b}}
 { e^{2\pi i\phi_a} - e^{2\pi i \phi_b}}\right\}^{h-1}
\end{eqnarray}
where $\mu_q(\phi)$ is defined in (\ref{one-loop det}).
Also, $\{\rm Sol\}$ is the set which satisfies (\ref{localization configuration}) and $0\le \phi_1 <\cdots < \phi_N<1$ as defined in section \ref{subsec:GWZWM localization}.

Let us show a correspondence between this one-point function and the expectation value of a conserved charge in the $q$-boson model. 
Before we study the correspondence, we prepare necessary knowledge for the $q$-boson model.
In particular, we show that the  expectation value of conserved charges in the $q$-boson model is expressed by power sums.
Recall that the eigenvalue of the transfer matrix (\ref{transfer matrix eigenvalue}) is expressed by 
\begin{eqnarray}
\Lambda(\mu,\{\lambda\}) 
= \prod_{j=1}^M \frac{\mu t-\lambda_j}{\mu-\lambda_j}
+\mu^L \prod_{j=1}^M \frac{\mu -\lambda_j t}{\mu-\lambda_j}.
\label{transfer matrix eigenvalue2}
\end{eqnarray}
The conserved charges $H_1,\cdots, H_L$ are given by expanding the transfer matrix $\tau(\mu)$ in terms of $\mu$ as explained in section \ref{subsec:q-boson ABA}:
\begin{eqnarray}
\tau(\mu) = \sum_{r=0}^L H_r \mu^r.
\end{eqnarray}
where $H_0=H_L=1$.
Therefore, we have
\begin{eqnarray}
H_r |\psi(\{\lambda\}_M)\rangle = \Lambda_r(\{\lambda\};t)|\psi(\{\lambda\}_M)\rangle
\end{eqnarray}
where $\Lambda_r(\{\lambda\};t)$ is the eigenvalue of the conserved charges  defined by
\begin{eqnarray}
\Lambda(\mu,\{\lambda\})
= \sum_{r=0}^L \Lambda_r(\{\lambda\};t)\mu^r .
\end{eqnarray}
Also, we have used the notation of  the Bethe state  defined in (\ref{Bethe state}).

Let us consider the eigenvalues of the conserved charges.
In order to do this, we define \cite{Macdonald:book}
\begin{eqnarray}
q_0(\lambda_1,\cdots,\lambda_M) &=& 1,\nonumber\\ 
q_r(\lambda_1,\cdots,\lambda_M) &=& (1-t) \sum_{j=1}^M \lambda_j^r \prod_{\substack{k=1\\k\neq j}}^M \frac{\lambda_j - t \lambda_k}{\lambda_j - \lambda_k}\quad{\rm for}\quad r\ge 1.
\end{eqnarray}
If $\{\lambda_1,\cdots,\lambda_M \}$ is the solution of the Bethe Ansatz equations (\ref{Bethe equation}), $q_r(\lambda_1,\cdots,\lambda_M ;t)$ satisfies 
\begin{eqnarray}
&& q_L(\lambda_1,\cdots,\lambda_M ;t) = 1 - t^M, \label{q_r constraint1}\\
&& q_L(\lambda_1,\cdots,\lambda_M^{-1} ;t) + t^M q_0(\lambda_1,\cdots,\lambda_M;t^{-1})= 1, \label{q_r constraint2}\\
&& q_{L+r}(\lambda_1,\cdots,\lambda_M;t) + t^M q_r(\lambda_1,\cdots,\lambda_M ; t^{-1})=0
\quad{\rm for}\quad r\ge 1.\label{q_r constraint3}
\end{eqnarray}
By using these relations, we rewrite the eigenvalue of the transfer matrix as follows:
\begin{eqnarray}
\Lambda(\mu,\{\lambda\})
= \sum_{r=0}^L \Lambda_r(\{\lambda\};t)\mu^r 
= \sum_{r=0}^{L-1} q_r(\lambda^{-1}_1\cdots,\lambda_M^{-1};t) \mu^r + \mu^L.
\end{eqnarray}
Then, we obtain the following expression for the eigenvalues of the conserved charges:
\begin{eqnarray}
\Lambda_r(\{\lambda\};t) &=& q_r(\lambda^{-1}_1\cdots,\lambda_M^{-1};t),\nonumber\\
\Lambda_L(\{\lambda\};t)&=& 1.
\end{eqnarray}
Moreover, we rewrite $q_r(\lambda_1,\cdots,\lambda_M ;t)$ as
\footnote{As we take the limit $t\rightarrow 0$, (\ref{q and power}) becomes a relation between a complete symmetric polynomial and   power sums:
\begin{eqnarray*}
h_r(\lambda_1,\cdots,\lambda_M;t)= \sum_{|\mu|=r}z_{\mu}^{-1}p_{\mu}(\lambda_1,\cdots,\lambda_M).
\end{eqnarray*}}
\begin{eqnarray}
q_r(\lambda_1,\cdots,\lambda_M;t)= \sum_{|\mu|=r}z_{\mu}(t)^{-1}p_{\mu}(\lambda_1,\cdots,\lambda_M)
\label{q and power}
\end{eqnarray}
where a power sum with partition $\mu$ is defined by
\begin{eqnarray}
p_{\mu}(\lambda_1,\cdots,\lambda_M) 
= p_{\mu_1}(\lambda_1,\cdots,\lambda_M) p_{\mu_2}(\lambda_1,\cdots,\lambda_M)\cdots p_{\mu_M}(\lambda_1,\cdots,\lambda_M).
\end{eqnarray}
and $z_{\mu}(t)$ is defined by 
\footnote{
Remark that we regard a partition with zero entries $\mu=(\mu_1,\cdots,\mu_s,0,\cdots,0)$ as 
$\mu=(\mu_1,\cdots,\mu_s)$ in (\ref{z def}).
For example,
\begin{eqnarray*}
z_{(2,0)}(t) \equiv z_{(2)}(t) = z_{(2)}\cdot (1-t^2).
\end{eqnarray*}}
\begin{eqnarray}
z_{\mu}(t) = z_{\mu}\cdot \prod_{j\ge 1}(1 - t^{\mu_j})
\quad{\rm and}\quad
z_{\lambda} = \prod_{i\ge 1} i^{m_i} m_i !.
\label{z def}
\end{eqnarray}
Then, we find that the eigenvalue of the conserved charges is expressed by the power sums with partitions by utilizing this relation as
\begin{eqnarray}
\Lambda_r(\{\lambda\};t) &=& \sum_{|\mu|=r}z_{\mu}(t)^{-1}p_{\mu}(\lambda_1^{-1},\cdots,\lambda_M^{-1}),\nonumber\\
\Lambda_L(\{\lambda\};t) &=&1.
\end{eqnarray}
Therefore, we obtain
\begin{eqnarray}
\frac{\langle \psi(\{\lambda\}_N)|H_r|\psi(\{\lambda\}_M)\rangle}{\langle \psi(\{\lambda\}_N)|\psi(\{\lambda\}_M)\rangle}
&=& \Lambda_r(\{\lambda\};t)\langle \psi(\{\lambda\}_N)|\psi(\{\lambda\}_M)\rangle\nonumber\\
&=& \sum_{|\mu|=r}z_{\mu}(t)^{-1}p_{\mu}(\lambda_1^{-1},\cdots,\lambda_M^{-1}).
\label{expectation values of conserved charge}
\end{eqnarray}

From now on, let us show  the correspondence between the one-point functions in the $SU(N)/SU(N)$ gauged WZW-matter model and the expectation values of the conserved charges in the $q$-boson model.
We first identify the level $k$, the dual Coxeter number $N$ of  $\mathfrak{su}(N)$, the coupling constant $\zeta$ and the Cartan part $\{\phi_1,\cdots,\phi_N\}$ of the field $g$  in the $SU(N)/SU(N)$ gauged WZW-matter model
with the total site number $L$, the total particle number $M$, the coupling constant $\eta$ and the Bethe roots $\{x_1,\cdots,x_N\}$ in the $q$-boson model, respectively,  as with the partition function.
Then, the one-point function can be expressed by the norms between the eigenstates in the $q$-boson model as follows:
\begin{eqnarray}
\hspace{-1cm}&&\langle \Tr g^r \rangle(\Sigma_h,SU(N),k, t)\nonumber\\
&&\hspace{0.5cm}= \left(\frac{N}{k}\right)^h \sum_{x_1,\cdots,x_N\in\{{\rm Sol}\}}\sum_{c=1}^Ne^{2\pi i r  \phi_c}
\langle \psi(\{ e^{2\pi ix}\}_N)|\psi(\{e^{2\pi ix}\}_N)\rangle^{h-1}.
\end{eqnarray}
We define special operators as 
\begin{eqnarray}
{\cal O}_r = \sum_{|\mu| = r} z_{\mu}(t)^{-1}\prod_{j=1}^N \Tr g^{\mu_j}
\end{eqnarray}
where $\mu = (\mu_1,\cdots,\mu_M)$ is a partition and $z_{\mu}(t)$ is defined in (\ref{z def}).
Then, we find that the one-point function of this operator is given by
\begin{eqnarray}
&&\hspace{-1cm}\langle {\cal O}_r  \rangle(\Sigma_h,SU(N),k, t) \nonumber\\
&=&   \left(\frac{N}{k}\right)^h \sum_{x_1,\cdots,x_N\in\{{\rm Sol}\}}
\Lambda_r(\{e^{2\pi i x}\};t)\cdot
\langle \psi(\{ e^{2\pi ix}\}_N)|\psi(\{e^{2\pi ix}\}_N)\rangle^{h-1}\nonumber\\
&=& \left(\frac{N}{k}\right)^h \sum_{x_1,\cdots,x_N\in\{{\rm Sol}\}}
\frac{\langle \psi(\{ e^{2\pi ix}\}_N)|H_r|\psi(\{e^{2\pi ix}\}_N)\rangle}{\langle \psi(\{ e^{2\pi ix}\}_N)|\psi(\{e^{2\pi ix}\}_N)\rangle}\nonumber\\
&&\vspace{5mm}
\hspace{3.8cm}\times \langle \psi(\{ e^{2\pi ix}\}_N)|\psi(\{e^{2\pi ix}\}_N)\rangle^{h-1}
\end{eqnarray}
where we have used (\ref{expectation values of conserved charge}).

As a result, we  have clarified the relations between the one-point functions of the special operators in the $SU(N)/SU(N)$ gauged WZW-matter model and the expectation values of the conserved charges in the $q$-boson model.
This also implies that the special operators ${\cal O}_r$ in the gauged WZW-matter model formally correspond to the conserved charges in the $q$-boson model.

The generalization to the $n$-point functions of ${\cal O}_{r_1},\cdots,{\cal O}_{r_n}$ is straightforward.
The $n$-point functions are given by
\begin{eqnarray}
&&\hspace{-1cm}\langle {\cal O}_{r_1} {\cal O}_{r_2} \cdots {\cal O}_{r_n} \rangle(\Sigma_h,SU(N),k, t)\nonumber\\
&=& \left(\frac{N}{k}\right)^h \sum_{x_1,\cdots,x_N\in\{{\rm Sol}\}}
\frac{\langle \psi(\{ e^{2\pi ix}\}_N)|H_{r_1}H_{r_2}\cdots H_{r_n}|\psi(\{e^{2\pi ix}\}_N)\rangle}
{\langle \psi(\{ e^{2\pi ix}\}_N)|\psi(\{e^{2\pi ix}\}_N)\rangle}\nonumber\\
&&\vspace{5mm}
\hspace{3.8cm}\times \langle \psi(\{ e^{2\pi ix}\}_N)|\psi(\{e^{2\pi ix}\}_N)\rangle^{h-1}.
\end{eqnarray}
Consequently,
we find that the Gauge/Bethe correspondence  works well for not only the partition function but also the correlation functions in  the topological gauge theory.

\section{Summary and Discussion}
In this paper, we have introduced a one-parameter deformation of the $G/G$ gauged WZW model by coupling it to BRST-exact matters 
and evaluated its partition function in the case of $G=SU(N)$ and $U(N)$.
We have shown that the localized field configurations in the path integral coincide with the Bethe Ansatz equations for the $q$-boson model and
 that the partition function  is represented by a summation of  the norms between the eigenstates in the $q$-boson model.
Thus, we have established the correspondence between the $U(N)/U(N)$ or $SU(N)/SU(N)$ gauged WZW-matter model and the $q$-boson model, which is a new example of the Gauge/Bethe correspondence.
This correspondence is a one-parameter deformation of ``Gauged WZW model/Phase model correspondence" \cite{Okuda:2012nx}.

We also have evaluated numerically the partition function and have given the explicit forms as the function of a deformation parameter $t$ in several cases.
This conjectured form of the partition function can be reproduced from the viewpoint of the axiom of the topological quantum field theory.
Then, we have shown that the $SU(N)/SU(N)$ gauged WZW-matter model is a lagrangian realization of a topological quantum field theory constructed by Korff \cite{Korff:2013rsa}.
This implies that the Gauge/Bethe correspondence is realized only if one constructs  a commutative Frobenius algebra from a certain integrable system, 
 as with \cite{Korff:2010, Korff:2013rsa}.
This is one of the reasons why the Gauge/Bethe correspondence works well.
Moreover, we have shown that the correlation functions can also be expressed by the language of  the $q$-boson model.
This is the correspondence for the quantity of a new type in the Gauge/Bethe correspondence.

We comment on several future directions.
We are interested in whether  the $G/G$ gauged WZW-matter model maintains  similar properties with the $G/G$ gauged WZW model.
First, we are interested in the Chern-Simons theory related to the $G/G$ gauged WZW-matter model. 
It is well known that the partition function of the $G/G$ gauged WZW model on $\Sigma_h$ coincides with that of the Chern-Simons theory with a gauge group $G$ on $S^1 \times \Sigma_h$. 
Therefore, we conjecture that there exists the three dimensional counterpart of the $G/G$ gauged WZW-matter model.
Since the gauged WZW-matter model possesses the scalar BRST charge which is crucial to carry out localization,
the three dimensional counterpart should  possess this property.
A natural candidate with a scalar BRST charge  is a (topologically) twisted supersymmetric Chern-Simons-matter theory on $S^1 \times \Sigma_h$.
 For example, a twisted Chern-Simons-matter theory on Seifert manifolds are constructed in \cite{Ohta:2012ev}.
 When we consider  the Chern-Simons theory coupled to an  adjoint twisted matter with a real mass $m$,  
 the one-loop determinant for this theory will coincide with (\ref{1-loop CS}) under the identification $t=e^{m}$. 
 It also is interesting to study a correspondence between the gauged WZW-matter model and a twisted Chern-Simons-matter theory in detail.

Secondly, it is known that the partition function of the $G/G$ gauged WZW model coincides with a geometric index 
over the moduli space $\mathcal{M}$ of the stable holomorphic $G_{\C}$-bundles  on a Riemann surface \cite{Witten:1993xi}:
\begin{eqnarray}
Z_{\rm GWZW}^G (\Sigma_h, k)= \int \mathrm{Td}(\mathcal{M}) \mathrm{ch} (\mathcal{L}^{\otimes k})  
=\mathrm{dim} {H}^0 (\mathcal{M} ; \mathcal{L}^{\otimes k}).
\label{indexflat}
\end{eqnarray} 
In the large level limit, the partition function is asymptotic to the volume of the moduli space of a flat connection 
 \cite{Witten:1991we}
 \footnote{The moduli space of the flat connection is isomorphic to that of the stable holomorphic $G_{\C}$-bundles  on a Riemann surface.},
and the action therefore reduces to that of the BF-theory whose partition function gives the volume of the moduli space of  flat connection \cite{Witten:1991we, Witten:1992xu}. 

How about the case of  the $G/G$ gauged WZW-matter model?
In the large level limit, the partition function of the BF-theory coupled to adjoint matters is interpreted as the volume of a moduli space
\begin{eqnarray}
\widetilde{\mathcal{M}}=\Bigl\{ (A,\Phi)  \Big|  F+ i[\Phi, \Phi^{\dagger}] d \mu=0,\, {\partial}_{A} \Phi=0, \, \bar{\partial}_{A} \Phi^{\dagger}=0 \Bigr\} / \mathcal{G}
\label{moduliYMH}
\end{eqnarray}
where $\mathcal{G}$ is the gauge transformation group of $G$. 
We conjecture that the partition function of the $G/G$ gauged WZW-matter model
is related to a certain geometric index over the moduli space $\widetilde{\mathcal{M}}$.
Thus,  the integrality of the partition function may be interpreted as
the dimensions of cohomologies. 

If $\Phi$ is not a section of  $ \mathrm{End}(E)$ but  a element of  $ \Omega^{(1,0)}(\Sigma_h, \mathrm{End}(E) )$,
 (\ref{moduliYMH}) is the moduli space of the Hitchin's equation or the Higgs bundle \cite{Hitchin:1986vp}.
 It is shown in \cite{Teleman:2003}  that there exists a index over the moduli space of the Higgs  bundle for a one-parameter deformation of (\ref{indexflat}).
 See also \cite{Telemannote}.
In this case, the deformation parameter is the parameter of the power series  of bundles over the moduli space.  
In their calculation, the index is  expressed by the summation over the solutions of nonlinear equations like  the localized equations of our model.
For example, see  (4,2) in \cite{Telemannote} for  $G_{\C}=SL(2,\C)$.   
It will be interesting  to give the interpretation for the partition function of our model in term of   a geometric index over $\widetilde{\mathcal{M}}$.

\subsection*{Acknowledgments}
We are grateful to  Tetsuo~Deguchi, Tohru~Eguchi, Saburo~Kakei,  Kazutoshi~Ohta, Tetsuya~Onogi, Norisuke~Sakai and Satoshi~Yamaguchi  for useful discussions and comments.



\begin{thebibliography}{10}
\baselineskip=15pt

\bibitem{Moore:1997dj}
  G.~W.~Moore, N.~Nekrasov and S.~Shatashvili,
  ``Integrating over Higgs branches,''
  Commun.\ Math.\ Phys.\  {\bf 209}, 97 (2000)
  [arXiv:hep-th/9712241].
  

\bibitem{Gerasimov:2006zt}
  A.~A.~Gerasimov and S.~L.~Shatashvili,
  ``Higgs bundles, gauge theories and quantum groups,''
  Commun.\ Math.\ Phys.\  {\bf 277}, 323 (2008)
  [arXiv:hep-th/0609024].
  


\bibitem{Okuda:2012nx} 
  S.~Okuda and Y.~Yoshida,
  ``G/G gauged WZW model and Bethe Ansatz for the phase model,''
  JHEP {\bf 1211}, 146 (2012)
  [arXiv:1209.3800 [hep-th]].


\bibitem{Bogoliubov:1993}
N.~M.~Bogoliubov, R.~K.~Bullough and G.~D.~Pang,
``Exact solution of a $q$-boson hopping model,''
  Phys.\ Lett.\ B {\bf 47} 11495 (1993).



\bibitem{Witten:1988hf}
  E.~Witten,
  ``Quantum field theory and the Jones polynomial,''
  Commun.\ Math.\ Phys.\  {\bf 121}, 351 (1989).
  
\bibitem{Blau:1993tv}
  M.~Blau and G.~Thompson,
  ``Derivation of the Verlinde formula from Chern-Simons theory and the G/G
  model,''
  Nucl.\ Phys.\  B {\bf 408}, 345 (1993)
  [arXiv:hep-th/9305010].

\bibitem{Nekrasov:2009ui}
  N.~A.~Nekrasov and S.~L.~Shatashvili,
  ``Quantum integrability and supersymmetric vacua,''
  Prog.\ Theor.\ Phys.\ Suppl.\  {\bf 177}, 105 (2009)
  [arXiv:0901.4748 [hep-th]].
  
\bibitem{Nekrasov:2009uh}
  N.~A.~Nekrasov and S.~L.~Shatashvili,
  ``Supersymmetric vacua and Bethe ansatz,''
  Nucl.\ Phys.\ Proc.\ Suppl.\  {\bf 192-193}, 91 (2009)
  [arXiv:0901.4744 [hep-th]].
 
 
\bibitem{Bogoliubov:1997}
N.~M.~Bogoliubov, A.~G.~Izergin and N.~A.~Kitanine,
``Correlation functions for a strongly correlated boson system,''
  Nucl.\ Phys.\ B {\bf 516}, 501 (1998)
  [arXiv:solv-int/9710002] 


 
 
  \bibitem{Atiyah:1989vu} 
  M.~Atiyah,
  ``Topological quantum field theories,''
  Inst.\ Hautes Etudes Sci.\ Publ.\ Math.\  {\bf 68}, 175 (1989).

\bibitem{Segal:2002ei} 
  G.~Segal,
  ``The definition of conformal field theory''.
  
  
\bibitem{Dijkgraaf:1997ip} 
  R.~Dijkgraaf,
  ``Les Houches lectures on fields, strings and duality,''
  In *Les Houches 1995, Quantum symmetries* 3-147
  [hep-th/9703136].

  \bibitem{Dijkgraaf}
    R.~Dijkgraaf,
  ``A Geometrical Approach to Two-Dimensional Conformal Field Theory,''
  Ph.D.Thesis (Utrecht, 1989).
 

\bibitem{Korff:2013rsa} 
  C.~Korff,
  ``Cylindric versions of specialised macdonald functions and a deformed Verlinde algebra,''
  Commun.\ Math.\ Phys.\  {\bf 318}, 173 (2013)
  [arXiv:1110.6356 [math-ph]].
 
     
\bibitem{Korff:2010}
C.~Korff and C.~Stroppel,
``The sl(n)-WZNW Fusion Ring: a combinatorial construction and a realisation as quotient of quantum cohomology,''
  Adv.\ Math.\ {\bf 225} (2010): 200-268
  [arXiv:0909.2347] .
 

  \bibitem{Korepin:book} 
V.~E.~Korepin, N.~M.~Bogoliubov and A.~G.~Izergin, 
``Quantum inverse scattering method and correlation functions,'' 
Cambridge University Press, Cambridge, 1993. 


\bibitem{Bethe:1931hc} 
  H.~Bethe,
  ``On the theory of metals. 1. Eigenvalues and eigenfunctions for the linear atomic chain,''
  Z.\ Phys.\  {\bf 71}, 205 (1931).

\bibitem{Yang:1968rm} 
  C.~N.~Yang and C.~P.~Yang,
  ``Thermodynamics of one-dimensional system of bosons with repulsive delta function interaction,''
  J.\ Math.\ Phys.\  {\bf 10}, 1115 (1969).

\bibitem{Slavnov:1989}
N.~A.~Slavnov, 
 ``Calculation of scalar products of wave functions and form factors in
the framework of the algebraic Bethe ansatz,"
 Theor.\ Math.\ Phys.\ {\bf 79},  (1989) 502-508. 

\bibitem{Slavnov:2007} 
   N.~A.~Slavnov,
  ``The algebraic Bethe ansatz and quantum integrable systems,''
  Russian Mathematical Surveys {\bf 62} (2007) 727.     


 
\bibitem{Deguchi:2009zz} 
  T.~Deguchi and C.~Matsui,
  ``Form factors of integrable higher-spin XXZ chains and the affine quantum-group symmetry,''
  Nucl.\ Phys.\ B {\bf 814}, 405 (2009)
  [Erratum-ibid.\ B {\bf 851}, 238 (2011)]
  [arXiv:0807.1847 [cond-mat.stat-mech]].
  
\bibitem{Deguchi:2010zz} 
  T.~Deguchi and C.~Matsui,
  ``Correlation functions of the integrable higher-spin XXX and XXZ spin chains through the fusion method,''
  Nucl.\ Phys.\ B {\bf 831}, 359 (2010)
  [arXiv:0907.0582 [cond-mat.stat-mech]].
  
\bibitem{Kulish:1991bk} 
  P.~P.~Kulish,
  ``Contraction of quantum algebras and q-oscillators,''
  Theor.\ Math.\ Phys.\  {\bf 86}, 108 (1991)
  [Teor.\ Mat.\ Fiz.\  {\bf 86}, 157 (1991)].
 

  
  
\bibitem{Gerasimov:1993ws}
  A.~Gerasimov,
  ``Localization in GWZW and Verlinde formula,''
  arXiv:hep-th/9305090.
 
\bibitem{Miyake:2011yr}
  A.~Miyake, K.~Ohta and N.~Sakai,
  ``Volume of Moduli Space of Vortex Equations and Localization,''
  Prog.\ Theor.\ Phys.\  {\bf 126}, 637 (2012)
  [arXiv:1105.2087 [hep-th]].

\bibitem{Manton:1998kq} 
  N.~S.~Manton and S.~M.~Nasir,
  Commun.\ Math.\ Phys.\  {\bf 199}, 591 (1999)
  [hep-th/9807017].

\bibitem{Witten:1991mm} 
  E.~Witten,
  ``On Holomorphic factorization of WZW and coset models,''
  Commun.\ Math.\ Phys.\  {\bf 144}, 189 (1992).

  
  \bibitem{Kock:book}
  J.~Kock,
    ``Frobenius algebras and 2D Topological Quantum Field Theories,''
    Cambridge, Cambridge University Press, 1985.
    
  \bibitem{Macdonald:book}
   I.~G.~Macdonald,
  ``Symmetric functions and Hall polynomials,''  
  Oxford University Press, 1979.

\bibitem{Ohta:2012ev} 
  K.~Ohta and Y.~Yoshida,
  ``Non-Abelian Localization for Supersymmetric Yang-Mills-Chern-Simons Theories on Seifert Manifold,''
  Phys.\ Rev.\ D {\bf 86}, 105018 (2012)
  [arXiv:1205.0046 [hep-th]].
  
\bibitem{Witten:1993xi} 
  E.~Witten,
  ``The Verlinde algebra and the cohomology of the Grassmannian,''
  In *Cambridge 1993, Geometry, topology, and physics* 357-422
  [hep-th/9312104].
  
\bibitem{Witten:1991we}
  E.~Witten,
  ``On quantum gauge theories in two-dimensions,''
  Commun.\ Math.\ Phys.\  {\bf 141}, 153 (1991).
  
  
\bibitem{Witten:1992xu}
  E.~Witten,
  ``Two-dimensional gauge theories revisited,''
  J.\ Geom.\ Phys.\  {\bf 9} (1992) 303
  [arXiv:hep-th/9204083].
 

  
\bibitem{Hitchin:1986vp} 
  N.~J.~Hitchin,
  ``The Selfduality equations on a Riemann surface,''
  Proc.\ Lond.\ Math.\ Soc.\  {\bf 55}, 59 (1987).

\bibitem{Teleman:2003}  
C.~Teleman and C.~T.~Woodward,
    ``The Index formula for the moduli of G-bundles on a curve,''
    Ann.\ Math {\bf 170}, (2009) 495-527.
 [arXiv:math/0312154].   

\bibitem{Telemannote}  
C.~Teleman,
    ``Loop Groups and G-bundles on curves,''
 http://math.berkeley.edu/$\sim$ teleman/lectures.html.   





  
  
\end{thebibliography}
\end{document}